\documentclass[aps,pre,twocolumn,floatfix]{revtex4-1}

\usepackage{bm}
\usepackage{graphicx}
\usepackage{amsmath}

\begin{document}

\title{Dispersionless pulse transport in mass-spring chains:
        \\All possible perfect Newton's cradles}

\author{Ruggero Vaia}
\affiliation{Istituto dei Sistemi Complessi, Consiglio Nazionale delle Ricerche,
	I-50019 Sesto Fiorentino, Italy}
\affiliation{Istituto Nazionale di Fisica Nucleare, Sezione di Firenze,
	I-50019 Sesto Fiorentino, Italy}

\date{\today}

\begin{abstract}
A pulse traveling on a uniform nondissipative chain of $N$ masses connected by springs is soon destructured by dispersion. Here it is shown that a proper modulation of the masses and the elastic constants makes it possible to obtain a periodic dynamics and a perfect transmission of any kind of pulse between the chain ends, since the initial configuration evolves to its mirror image in the half period. This makes the chain to behave as a Newton's cradle. By a known algorithm based on orthogonal polynomials one can numerically solve the general inverse problem leading from the spectrum to the dynamical matrix and then to the corresponding mass-spring sequence, so yielding all possible ``perfect cradles''. As quantum linear systems obey the same dynamics of their classical counterparts, these results also apply to the quantum case: for instance, a wavefunction localized at one end would evolve to its mirror image at the opposite chain end.

\end{abstract}

\maketitle

\section{Introduction}
\label{s.intro}

\begin{figure}
	\includegraphics[width=0.47\textwidth]{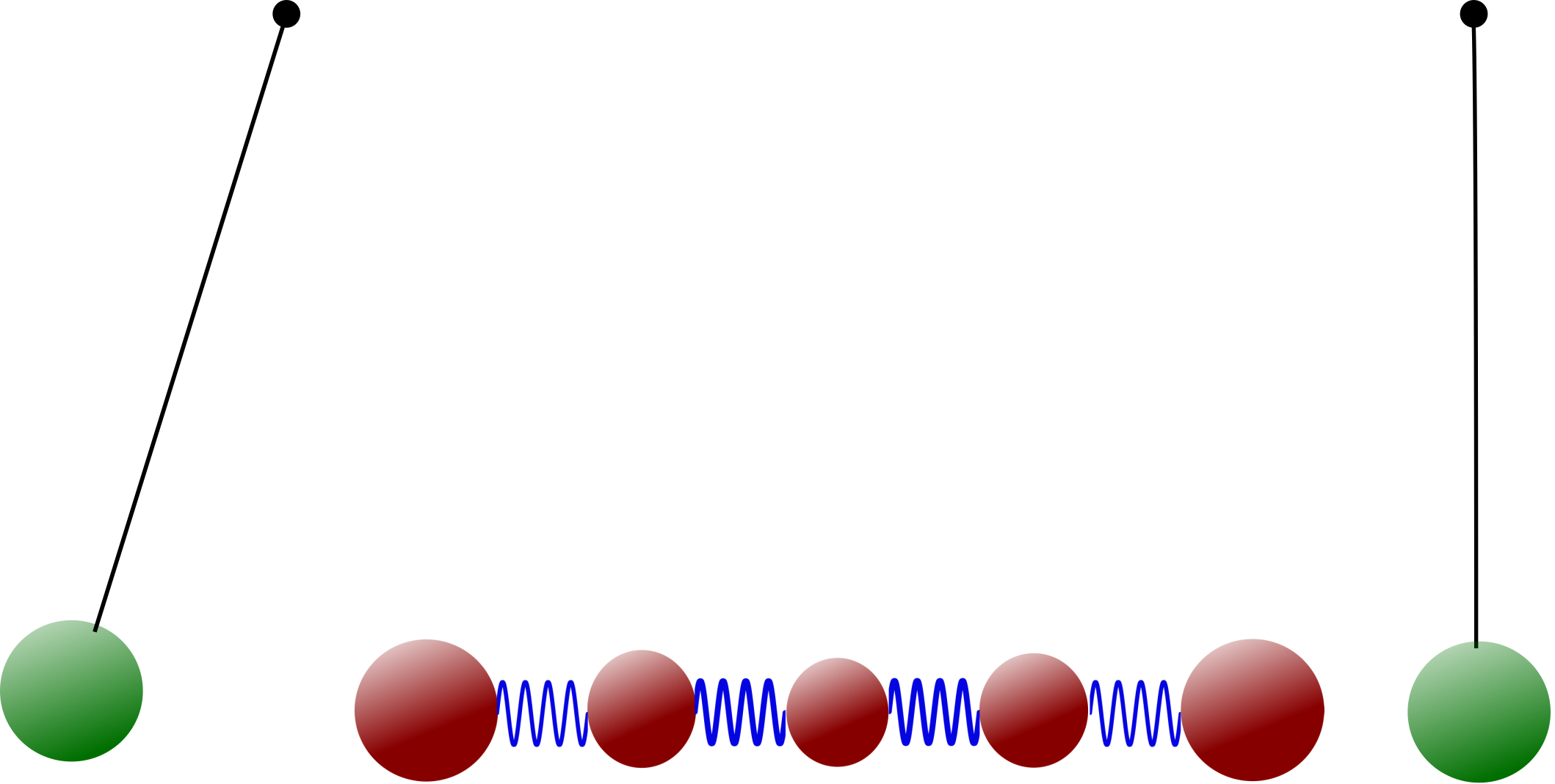}
	\caption{A perfect 5-mass chain. The masses are proportional to the sequence (35, 20, 18, 20, 35) and the springs' elastic constants to $(7, 9, 9, 7)$: this setup yields perfect end-to-end transmission~\cite{HerrmannS1981}. The auxiliary (green) external masses, equal to the first/last ones, behave just like in a Newton's cradle, except for the finite time between subsequent bounces. Chains of arbitrary length can behave in the same way, provided that the sequence of mass and spring values be properly chosen.}
	\label{f.cradle5}
\end{figure}

\begin{figure}
	\includegraphics[width=0.47\textwidth]{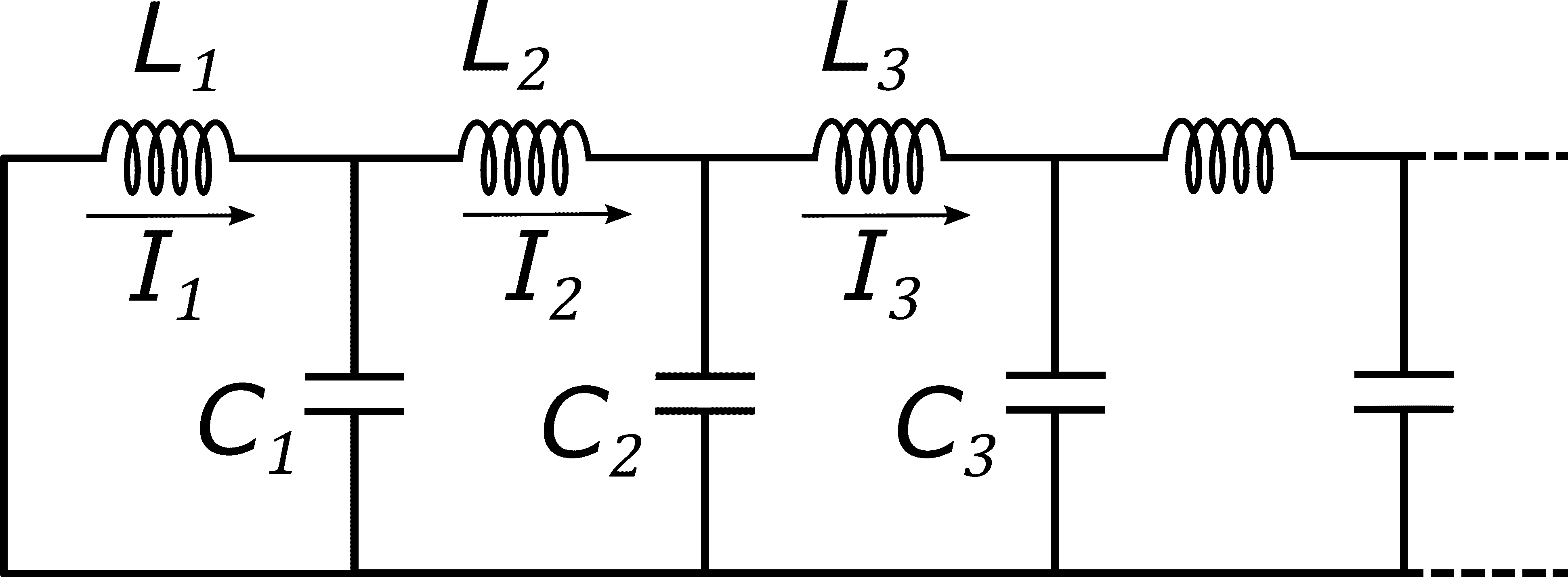}
\caption{An electric circuit equivalent to the spring-mass chain~\eqref{e.Horig}: the masses and the elastic constants are replaced by the inductances and inverse capacitances, respectively: $m_i=L_i$ and $K_i={C^{-1}_i}$. The charges $Q_i$ flowing in the conductances ($\partial_t{Q}_i=I_i$) play the role of the displacements and the momenta are $P_i=L_iI_i$. The charge on capacitor $C_i$ is $Q_i-Q_{i+1}$.}
	\label{f.LCcircuit}
\end{figure}

By mass-spring chain one means a sequence of $N$ masses $\{m_i\}$ connected by $N{-}1$ springs obeying Hooke's law and characterized by their elastic constants $\{K_i\}$. The system's Hamiltonian reads 
\begin{equation}
{\cal{H}} = \sum_{i=1}^N \frac{P_i^2}{2m_i}
+ \frac12\sum_{i=1}^{N-1}K_i(Q_i-Q_{i{+}1})^2 ~,
\label{e.Horig}
\end{equation}
where $Q_i$ is the displacement (measured from the equilibrium position) and $P_i$ the momentum of the $i$-th mass. This physical model, depicted in Fig.~\ref{f.cradle5}, is very general: for instance, it can be realized by the electric LC circuit of Fig.~\ref{f.LCcircuit}, where springs and masses are replaced by capacitors and inductors, respectively, while displacements are represented by the capacitors' charges.

Nanoscopic realizations of the model~\eqref{e.Horig} are relevant for their end-to-end transport properties~\cite{CahillEtAl2003,NorrisLB2013} and even atomic chains, which can be suitably described by the mass-spring chain model, have been created and characterized~\cite{RammPH2014}, e.g., studying the transmission of a momentum pulse given to an extremal atom.

The purpose of this paper is to demonstrate the existence, for any chain length $N$, of an infinite number of sequences $\{m_i\}$ and $\{K_i\}$ which can yield the perfect end-to-end transmission of such a pulse. For this reason such mass-spring chains behave in analogy to the popular Newton's cradle~\cite{HerrmannS1982,HutzlerDWM2004}, a mechanical device that displays an almost perfect transfer of momentum between the endpoints of an array of metallic spheres. Dubbing them ``mass-spring Newton's cradles'' is therefore natural, also looking at Fig.~\ref{f.cradle5}, and justified by the wide use of this terminology in physical systems where one can observe the perfect transfer of a localized conserved quantity over a one-dimensional structure, e.g., the ``quantum Newton's cradle'' of Ref.~\cite{KinoshitaWW2006}.

It is well known that a {\em uniform} mass-spring chain, made of identical masses $m_i=m$ and identical springs $K_i=K$, cannot efficiently transfer a pulse, even in the absence of dissipation, due to the effect of {\em dispersion}. Indeed, its normal modes, which coincide with the Fourier components of the coordinates and momenta, do have incommensurate frequencies~\cite{Kittel1953},
\begin{equation}
 \omega_n=2\sqrt{\frac{K}{m}}\,\sin\frac{\pi(n-1)}{2N},,~~~~n=1,...,N~,
\label{e.omegan.uniform}
\end{equation}
$n$ being the normal-mode label. It follows that their time evolution, ruled by the phase factors $e^{-i\omega_nt}$, can never lead to {\em coherently} recombining their amplitudes, i.e., with equal phases.

It is worth to mention that, historically, the research for coherent transmission of signals along one-dimensional systems preferred to renounce to the linearity (i.e., using non-Hooke springs), while keeping translation invariance, i.e., uniformity. Starting from the famous Fermi-Pasta-Ulam numerical simulation~\cite{FermiPU1955}, more and more studies enlightened the properties of nonlinear models and showed that nonlinearity can be an antidote against dispersion; several examples of localized and coherently propagating excitations, the ``solitons'', were discovered, such as in the {\em Toda lattice}, or in the continuous systems described by the {\em Korteweg-De Vries} equation (known since the 19th century) and the {\em Nonlinear Schr\"odinger} equation, just to mention the most well-known ones~\cite{DauxoisP2006}.

At variance with this way to reach coherence, in this paper Hooke's law is kept and translation invariance is instead abandoned, so dealing with {\em nonuniform} chains. Conceding a limited degree of nonuniformity, it was possible~\cite{Vaia2018} to enhance the transmission properties of the linear chain~\eqref{e.Horig} keeping it uniform in the bulk and symmetrically tuning two masses and their spring at both ends, in order to maximize the transmission of a pulse given to the first mass. Such a tuning has two main effects, namely, modulation of the amplitudes of the excited normal modes and deformation of the frequency spectrum: the best trade-off between them produces a huge improvement in the end-to-end transmission efficiency, attaining 98.7\,\% in the asymptotic limit of infinite length~\cite{Vaia2018}. The explanation is that only few normal modes with nearly equally-spaced frequencies are involved in the dynamics of the initial kick. However, this approach is very particular: it yields optimal, but not perfect, transmission and only for excitations localized on the first mass; a kick given, say, to the second mass would not coherently reach the opposite end.

Allowing for full nonuniformity, here a different questions are raised: Can one obtain ``magic'' mass-spring sequences that yield an exactly periodic and perfectly transmitting dynamics? How many of those sequences exist?

An obvious requirement for coherence is that the normal-mode frequencies be commensurate, i.e., integer multiples of a finite frequency $\omega$, say
\begin{equation}
 \omega_n=\omega\,{k_n},
\label{e.omegan.kn}
\end{equation}
with $\{k_n\}$ any sequence of integers (with no common factors, to be absorbed into $\omega$). Indeed, the dynamical phases, evolving as $e^{i\omega\,k_nt}$, would become unity after a time period $2\pi/\omega$ (and integer multiples of it); then, the normal modes would coherently recombine to exactly reproduce the initial configuration. It is again obvious that, in order to yield the spectrum~\eqref{e.omegan.kn}, one must renounce the assumption of uniform masses and springs along the chain, as they give~\eqref{e.omegan.uniform}. However, in order to be effective for transmission, the chain must allow for pulses traveling from one extremity to reproduce themselves at the opposite one without changes in shape: this entails that the chain has to be at least {\em mirror symmetric}, i.e., the transformation of inverting the sequence of masses and springs is a symmetry,
\begin{equation}
 m_i = m_{N{+}1-i}
~,~~~~~
 K_i = K_{N-i} ~;
\label{e.mKmirror}
\end{equation}
the $2N\,{-}\,1$ parameters $\{m_i,K_i\}$ are therefore reduced  to $N$ independent ones. With this assumption, it will be shown that if the above defined integers $\{k_n\}$ are alternating in parity, then at the half period $t^*=\pi/\omega$ (and odd multiples of it) the chain configuration becomes the mirror image of the initial one, so that any pulse at one end would be transferred to the opposite end with identical shape. It will be also proven that the ``magic'' mass-spring sequences exist and are in one-to-one correspondence with the distinct successions of the $N$ integers $\{k_n\}$ defined in Eq.~\eqref{e.omegan.kn}, attaining the goal of finding and classifying all perfect chains, i.e., those with full 100\,\% transmission efficiency whatever the shape of the initial pulse.

As for the mathematical methods, the task of finding the normal modes and the frequencies of the mass-spring chain~\eqref{e.Horig} can be reduced, by using mass-weighted canonical variables, to the diagonalization of a tridiagonal symmetric matrix, usually dubbed a {\em Jacobi} matrix~\cite{Parlett1998}. The chain's mirror symmetry further entails it to be symmetric also with respect to the second diagonal, so one deals with a {\em persymmetric} Jacobi matrix.
The goal pursued here is however the {\em inverse problem}, namely one starts from the desired eigenvalue succession~\eqref{e.omegan.kn} and wishes to obtain the corresponding Jacobi matrix and, in turn, the related mass-spring sequence. Fortunately, it is known that the inverse problem of calculating the elements of a persymmetric Jacobi matrix, such that its eigenvalues be a given nondegenerate~\cite{note1} sequence, is well-posed and the solution has been proven to exist~\cite{Hochstadt1967} and to be unique~\cite{Hald1976}. 
Moreover, the matrix elements can be calculated, at least numerically, by means of efficient algorithms~\cite{deBoorG1978}. The remaining task of relating these elements to the values of the masses and of the spring constants, can also be unambiguously solved~\cite{NylenU1997}.

Remarkably, an explicit analytic solution to the inverse problem was recently found~\cite{VaiaS2020} for any $N$ when $k_n=n-1$, i.e., the frequencies are proportional to the sequence of the first $N$ integers,
\begin{equation}
 \omega_n=\omega\,(n{-}1),~~~~n=1,...,N~.
\label{e.linear}
\end{equation}
For instance, with $N=5$ one gets the chain of Fig.~\ref{f.cradle5}. For $N\le5$ the exact solution was already known~\cite{HerrmannS1981}, though the calculations were not published, being ``somewhat lengthy''. 
On the other hand, for arbitrary sequences of integers in Eq.~\eqref{e.omegan.kn} a numerical approach cannot be avoided. It will be shown that a variety of perfect chains can be obtained, with some of them even more interesting as they entail smaller imbalances between largest and smallest masses and elastic constants.
All these ``perfect chains'' can be employed in different contexts for transmitting localized pulses (say, energy, heat, sound, etc.) between their extrema; perhaps such a mechanism might be already used by Nature, e.g., for transferring energy inside biological structures. It is to be emphasized that, since the model system is linear, the corresponding quantum model also shows an analogous behavior; for instance, the quantum wave function of such an array evolves in the time $t^*$ to its mirror-symmetric counterpart, which amounts to state that a localized wavepacket at one end would be perfectly transmitted to the opposite end of the chain. 

The paper is organized as follows. Basic notations and definitions are briefly recalled in Section~\ref{s.model}, as well as the dynamics of the chain in terms of its normal modes. In Section~\ref{s.transmission} the transmission amplitude is defined as indicator of the pulse-transfer efficiency. The inverse problem and the solution algorithm are the subject of Section~\ref{s.inverse}. Eventually, in Section~\ref{s.numerical} the dynamics of different mass-spring chains are compared and discussed, including the uniform, the optimized quasi-uniform~\cite{Vaia2018}, and the perfect chains with the spectrum~\eqref{e.linear} and more selected perfect chains.

\section{The free mass-spring chain}
\label{s.model}

Consider the chain described by the Hamiltonian~\eqref{e.Horig}; the absence of external springs ({\em free-free} boundary conditions, i.e., $K_0=K_N=0$) entails translation invariance, so that the system is expected to possess a zero-frequency normal mode (translation mode).
In terms of the displacement and momentum vectors, ${\bm{Q}}\equiv\{Q_i\}$ and  ${\bm{P}}\equiv\{P_i\}$, the same Hamiltonian can be written in matrix form,
\begin{equation}
{\cal{H}} = \frac12\, {\bm P}^{\rm T}\bm{M}^{-1}{\bm P}
   + \frac12\, {\bm Q}^{\rm T} {\bm K} {\bm Q} ~,
\end{equation}
where the ``mass matrix'' $\bm{M}$ is diagonal, its elements being $\{M_{ij}=m_i\,\delta_{ij}\}$, and
\begin{equation}
 \bm{K}=
 \begin{bmatrix}
	 K_1 & -K_1       &  0            &\cdots \\[2mm]
	-K_1 & K_1{+}K_2 & -K_2       &\cdots \\[2mm]
	 0      & -K_2       & K_2{+}K_3 &       \\[2mm]
	\vdots  & \vdots        &               &\ddots
\end{bmatrix}_{N} ~,
\label{e.K}
\end{equation}
is the symmetric tridiagonal ``elastic matrix''. Its rows sum up to zero, so that $\bm{K}$ has the eigenvector $(1,~1,~...,~1)$ corresponding to the translation mode with eigenvalue zero, so ${\rm{det}}\,{\bm{K}}=0$. The canonical transformation to mass-weighted coordinates, ${\bm{q}}={\bm{M}}^{1/2}{\bm{Q}}$ and ${\bm{p}}={\bm{M}}^{-1/2}{\bm{P}}$, turns the Hamiltonian into
\begin{equation}
 {\cal{H}} = \frac12\, {\bm p}^{\rm T}{\bm p}
    + \frac12\, {\bm q}^{\rm T} {\bm A} {\bm q} ~.
\label{e.H}
\end{equation}
The $N{\times}N$ matrix ${\bm{A}}={\bm{M}}^{-1/2}{\bm{K}}{\bm{M}}^{-1/2}$ is a {\em Jacobi} matrix, i.e., a tridiagonal symmetric matrix,
\begin{equation}
\bm{A} = 
\begin{bmatrix}
	 a_1   & -b_1 &  0    &\cdots    &  0      \\
	-b_1   &  a_2 & -b_2  &          & \vdots  \\
	 0     & -b_2 &  a_3  &\ddots    &  0      \\
	\vdots &      &\ddots &\ddots    & -b_{N-1}\\
	 0     &\cdots&  0    & -b_{N-1} &  a_N
\end{bmatrix}_{N}~,
\label{e.AN}
\end{equation}
whose nonzero elements are
\begin{align}
\begin{aligned}
 a_i &= \frac{K_{i-1}+K_i}{m_i}~,&~~~ &i=1,..., N~,
\\
 b_i &= \frac{K_i}{\sqrt{m_im_{i{+}1}}}~,& &i=1,\,...,\,N{-}1~.
\label{e.aibi}
\end{aligned}
\end{align}

The assumption of mirror symmetry~\eqref{e.mKmirror} entails that the matrices ${\bm{M}}$ and ${\bm{K}}$ are {\em persymmetric} (symmetric with respect to the antidiagonal), and the same holds for ${\bm{A}}$,
\begin{equation}
 a_i = a_{N{+}1-i}
~,~~~
 b_i = b_{N-i} ~.
\label{e.abmirror}
\end{equation}
Note that as ${\rm{det}}\,{\bm{A}}=0$ only $N{-}1$ of these matrix elements are independent. They are in correspondence with the $N$ independent parameters $\{m_i,K_i\}$, apart from an overall factor: indeed, Eq.~\eqref{e.aibi} shows that such a factor does not affect $\bm{A}$, reflecting the fact that scaling masses and spring constants by the same factor does not affect the system's frequencies and normal modes. This choice is arbitrary, e.g., one can fix the first mass $m_1$ or the total mass~\cite{GantmacherK1950}, and fully completes the mapping between $\bm{A}$ and the pair $(\bm{M},\bm{K})$

As all $b$'s are nonzero the eigenvalues of the matrix $\bm{A}$ are distinct~\cite{Parlett1998}; moreover, they are nonnegative since  $\bm{A}$ is positive semi-definite,
\begin{equation}
 \sum_{ij}A_{ij}q_iq_j=\sum_iK_i(Q_i-Q_{i+1})^2\ge0~, ~~~~ \forall \{q_i\} ~.
\end{equation}
Denoting by $\bm{U}\,{=}\,\{U_{ni}\}$ the orthogonal matrix that diagonalizes $\bm{A}$,
\begin{equation}
 \sum_{ij}U_{ni}A_{ij}U_{mj} = \lambda_n\,\delta_{nm} ~,
\label{e.Uni}
\end{equation}
and introducing the normal-mode coordinates and momenta,
\begin{equation}
 \tilde{q}_n=\sum_{i=1}^NU_{ni}\,q_i
~,~~~
 \tilde{p}_n=\sum_{i=1}^NU_{ni}\,p_i ~,
\label{e.pnqn}
\end{equation}
the Hamiltonian~\eqref{e.H} becomes a sum of independent oscillators,
\begin{equation}
 {\cal{H}}=\frac12\sum_{n=1}^N \big(\tilde{p}_n^2 + \omega_n^2\,\tilde{q}_n^2\big) ~.
\end{equation}
where the eigenfrequencies are the positive square roots of the eigenvalues,
$\omega_n=\sqrt{\lambda_n}$. They are assumed in increasing order, $\omega_{n+1}>\omega_n$, starting from $\omega_1=0$.

The chain's time evolution is a superposition of normal-mode motions,
\begin{equation}
 q_i(t) =  \sum_{n=1}^N U_{ni}\sum_{j=1}^N U_{nj}\,\Big[q_j(0)\cos\omega_nt +{p_j(0)}\frac{\sin\omega_nt}{\omega_n}\Big] ~.
\label{e.solution}
\end{equation}
All frequencies are positive, except that of the translation mode, $\omega_1=0$: its corresponding component is to be understood as the overall translation $U_{1i}\sum_jU_{1j}\big[q_j(0)+p_j(0)\,t\big]$.

\section{Perfect pulse transmission}
\label{s.transmission}

The transmission of a pulse between the chain ends can be described as follows. Assume the first mass is given an instantaneous kick, i.e., a given momentum $\bar{p}$, as in experiments with ion chains~\cite{RammPH2014},
\begin{equation} 
 \bm{q}(0)=(0,\,0,\,...,\,0)
~,~~~~ \bm{p}(0)=(\bar{p},\,0,\,...,\,0) ~.
\label{e.q0p0}
\end{equation}
One seeks for values of the chain parameters~\eqref{e.mKmirror} such that the dynamics leads in a certain time $\tau$ as close as possible to the mirror-symmetric momentum distribution $\bm{p}(\tau)=(0,\,0,\,0,\,...,\,\bar{p})$. With these initial conditions Eq.~\eqref{e.solution} gives
\begin{equation}
  p_i(t) =\partial_tq_i(t) = \bar{p}\sum_{n=1}^NU_{ni}U_{n1}\cos\omega_nt ~.
\label{e.qit}
\end{equation}
The {\em transmission amplitude} $\alpha_{_N}(t)\equiv{p_{_N}(t)}/{\bar{p}}$ is defined as the ratio between the momentum of the last mass at time $t$ and the input momentum of the first mass. Using the fact that the eigenvectors of $\bm{A}$ alternate between mirror-symmetric and -antisymmetric~\cite{CantoniB1976,Parlett1998}, $U_{n,N+1-i}\,{=}\,(-)^{n-1}U_{ni}$, one has
\begin{equation}
 \alpha_{_N}(t) = \sum_{n=1}^NU_{n1}^2\cos[\pi(n{-}1){-}\omega_nt] ~.
\label{e.utn}
\end{equation}
The numbers $U_{n1}^2$ weigh the contributions from the normal modes and can be regarded as a normalized probability density, since $\sum_nU_{n1}^2\,{=}\,1$.
The same parameter $\alpha_{_N}(t)$ characterizes the transmission of an initial elongation of the first mass, ${\bm{q}}(0)=(\bar{q},\,0,\,...,\,0)$, while the chain is at rest, $\bm{p}(0)=\bm{0}$: indeed, Eq.~\eqref{e.solution} yields $q_{_N}(t)=\bar{q}\,\alpha_{_N}(t)$.

Perfect transmission occurs when at some time instant $t^*$ all phases are coherent, i.e., they are equal or differ by integer multiples of $2\pi$:
\begin{equation}
 \pi(n{-}1)-\omega_nt^*=\pi\times~{\rm even~integer} ~.
\end{equation}
This amounts to require that the frequencies are integer multiples of a characteristic frequency $\omega=\pi/t^*$,
\begin{equation}
 \omega_n=\omega\,k_n ~,
\label{e.om_kn}
\end{equation}
with the (different) coprime integers $k_n$ having the same parity of $n{-}1$, since $k_1=0$; it is equivalent to require that the increments
\begin{equation}
 \delta_n \equiv k_{n+1}-k_n ~,~~~ i=1,\,...,\,N{-}1~,
\label{e.deltan}
\end{equation}
be {\em odd positive integers} with no common factors (i.e., coprime). In the next section it is shown that that any given frequency spectrum corresponds to a particular mass-spring sequence, so it follows that for any $N\ge3$ there is a countably infinite set of different chains yielding perfect transmission.

At time $2t^*$ the chain returns to the initial state, as, e.g., $p_1(2t^*)=\bar{p}\sum_{n=1}^NU_{n1}^2\cos(2\pi\,k_n)=\bar{p}$, showing a perfectly periodic dynamics~\cite{note2}. The time evolution consists in the propagation of the initial pulse along the chain, with a shape involving displacements of all masses. Actually, perfect behavior does not require the particular initial configuration~\eqref{e.q0p0}: Eq.~\eqref{e.solution} with the spectrum~\eqref{e.om_kn}-\eqref{e.deltan} tells that any initial shape of the chain evolves to its exact mirror image at $t^*$ and is restored at $2t^*$. Such a perfect mass-spring cradle is depicted in Fig.~\ref{f.cradle5}: to make it resemble a Newton's cradle, it is imagined to involve two auxiliary hanging masses that periodically transmit/receive momentum by instantaneous hard-sphere collision with the chain extrema.

The upcoming sections are devoted to the calculation of the ``magic'' mass-spring sequences that determine $\alpha_N(t^*)=1$, i.e., 100\,\% transmission amplitude for any value of $N$.

\section{The inverse problem}
\label{s.inverse}

The task of finding the matrix elements of the tridiagonal symmetric and mirror-symmetric matrix~\eqref{e.AN} starting from the requirement that it have a given spectrum $\{\lambda_n,~n=1,\,...,\,N\}$ is an ``inverse problem''. It is well-posed, as in the case considered here (free-free boundary conditions) the number of independent matrix elements to be determined, $N{-}1$, is equal to the number of input variables, the positive eigenvalues.

In the previous Section it was shown that the necessary and sufficient condition to yield perfect transmission is that the mode frequencies be given by Eq.~\eqref{e.om_kn}, where the increasing sequence of integers $\{k_n\}$ obeys the constraint~\eqref{e.deltan}.

\subsection{Simplest case: analytic solution}

The simplest choice for  the odd-number sequence~\eqref{e.deltan} is $\delta_n=1$, or $k_n=n\,{-}\,1$, corresponding to the spectrum~\eqref{e.linear}. In this case an analytic solution was recently obtained~\cite{VaiaS2020}, based on the following result, whose  proof is sketched in Appendix~\ref{a.proof}. Let $\bm{A}$ be the $N{\times}N$ matrix~\eqref{e.AN} with the mirror-symmetric entries
\begin{equation}
\begin{aligned}
 a_i&=N{-}1+4(i{-}1)(N{-}i) ~, &&i=1,\,...,\,N~,
\\
 b_i&=\sqrt{i\,(2i{-}1)\,(N{-}i)\,(2N{-}1{-}2i)} ~,&&i=1,\,...,\,N{-}1~;
\label{e.ex.aibi}
\end{aligned}
\end{equation}
then its eigenvalues are
\begin{equation}
 \lambda_n=2(n{-}1)^2 ~,~~~~~~n=1,\,...,\,N ~.
\label{e.lambdaN}
\end{equation}
Both $a_i$ and $b_i$ are of order $N^2$ in the matrix bulk, and decrease almost parabolically toward the matrix edges, where they are of order $N$. Their imbalance, namely the ratio between largest and smallest entries, is of order $N$.

\begin{figure}
\includegraphics[width=0.47\textwidth]{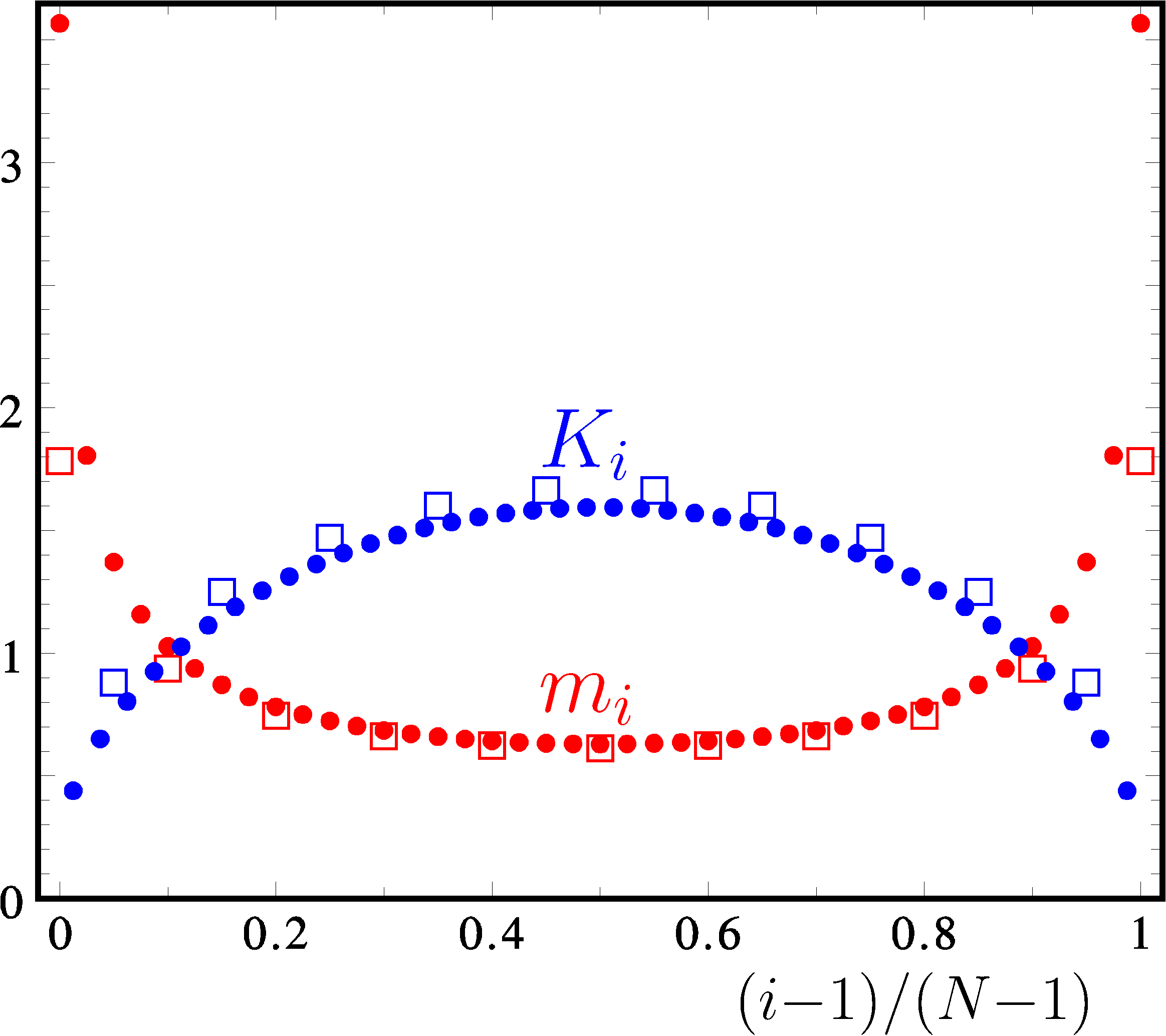}
\caption{The sequence of masses $\{m_i\}$ and elastic constants $\{K_i\}$ as given by Eqs.~\eqref{e.MiKi} for chains of size $N=11$ (squares) and $N=41$ (bullets). Choosing $m_1=\sqrt{(N{-}1)/\pi}$, $\omega=\pi/(N{-}1)$, and the scaled variable $(i{-}1)/(N{-}1)$ allows to appreciate the overall scaling behavior~\cite{VaiaS2020}. As the springs $K_i$ connect $m_i$ and $m_{i+1}$, for clarity they are plotted at $i{+}\frac12$.}
	\label{f.miki}
\end{figure}

Hence, the frequency sequence $\omega_n=\omega\,(n\,{-}\,1)$ can be obtained by imposing a factor $\omega^2/2$ to the expressions~\eqref{e.ex.aibi}.
The sequences of masses and elastic constants that produce the interaction matrix $\bm{A}$ through the transformation~\eqref{e.aibi} admit closed expressions~\cite{VaiaS2020} in terms of binomial coefficients,
\begin{equation}
\begin{aligned}
 m_i &= m_1\binom{N{-}1}{i{-}1}^2 ~\binom{2N{-}2}{2i{-}2}^{-1}
\\
 K_i &= m_1\omega^2(N{-}1)^2~\binom{N{-}2}{i{-}1}^2 ~\binom{2N{-}2}{2i{-}1}^{-1} ~.
\end{aligned}
\label{e.MiKi}
\end{equation}
It turns out that for $i{+}1<n/2$ it is
\begin{equation}
 M_{i+1}<M_i ~, ~~~~~~~~ K_{i+1}>K_i ~,
\end{equation}
implying that the smallest masses and the largest elastic constants lie in the middle of the chain. For large $n$ one finds that the ratio between largest and smallest values is of order $\sqrt{N}$. The binomials being rational numbers, one can choose $m_1$ and $\omega$ in such a way that all $\{m_i\}$ and $\{K_i\}$ be expressed by coprime integers: a few of these ``magic numbers'' are shown in Table~1 of Ref.~\cite{VaiaS2020}. The sequences of masses and elastic constants are graphically reported in Fig.~\ref{f.miki} for $N=11$ and $N=41$.

\subsection{General case: numerical solution}

For a general choice of the eigenvalues~\eqref{e.om_kn}, the inverse problem has to be faced numerically. To this purpose a good algorithm was proposed by de\,Boor and Golub~\cite{deBoorG1978} (BG). It constructs the sequence of characteristic polynomials $\{\chi_i(\lambda)\}$~ $(i=0,\,...,\,N)$ of the matrix $\bm{A}$ and its submatrices, using their orthogonality with respect to the internal product
\begin{equation}
 \big\langle \chi, \tilde\chi \big\rangle 
 \equiv \sum_{n=1}^N w_n~ \chi(\lambda_n)~\tilde\chi(\lambda_n) ~,
\end{equation}
where the weights are defined in terms of the required eigenvalues,
\begin{equation}
 w_n=w\prod_{\substack{m=1\\m\ne{n}}}^N\big|\lambda_n-\lambda_m\big|^{-1} ~,
\label{e.w}
\end{equation}
$w$ being an arbitrary positive constant.
The property of orthogonality allows for the sequential construction of the polynomials,
\begin{equation}
 \chi_{i+1}(\lambda)=(\lambda-a_{i+1})\,\chi_i(\lambda)-b_i^2\chi_{i-1}(\lambda)~,
\label{e.chii}
\end{equation}
starting from $\chi_0(\lambda)=1$, $b_0(\lambda)\equiv{0}$, with the coefficients
\begin{equation}
 a_{i+1}=\frac{\big\langle \lambda\,\chi_i, \chi_i\big\rangle}
                 {\big\langle \,\chi_i, \chi_i\big\rangle}
~,~~~~~
 b_i^2=\frac{\big\langle \,\chi_i, \chi_i\big\rangle}
           {\big\langle \,\chi_{i-1}, \chi_{i-1}\big\rangle}
\label{e.ai_bi}
\end{equation}
corresponding to the matrix elements one is looking for.

The numerical procedure starts by calculating and storing the weights~\eqref{e.w} and proceeds in a simple manner with the iteration of Eqs.~\eqref{e.chii} and~\eqref{e.ai_bi}. It takes few lines of code, a clear example being found in Ref.~\cite{BrudererFRBO2012}.

Once the matrix $\bm{A}$ is known, one has to reconstruct the sequence of masses and springs~\cite{NylenU1997}. The masses are related to the components of the translation-mode eigenvector, $\bm{v}=\{U_{1i}\}$, since the identity
\begin{equation}
 0=\bm{v}^t\bm{A}\,\bm{v} =\sum_{i=1}^{N-1}K_i
 \bigg(\frac{v_i}{\sqrt{m_i}}-\frac{v_{i+1}}{\sqrt{m_{i+1}}}\bigg)^2
\end{equation}
entails the ratios $v_i/\sqrt{m_i}$ to be equal, hence $m_i=c\,v_i^2$, with $c$ a constant that is determined by the choice of $m_1$. The components of the equation $\bm{A}\,\bm{v}=\bm{0}$ define a recursion relation,
\begin{equation}
 v_{i+1}=\frac{v_i\,a_i-v_{i-1}\,b_{i-1}}{b_i} ~,
\end{equation}
that can be used to obtain the mass sequence starting from the given first mass (see Sec.~\ref{s.model}); in this way one has $v_2=a_1v_1/b_1$, then $v_3=(a_1a_2-v_1b_1^2)/(b_1b_2)$, and so on. Eventually, the elastic constants follow from Eq.~\eqref{e.aibi}, $K_i=c\,v_iv_{i+1}\,b_i$.

\section{Results and comparisons}
\label{s.numerical}

The aim of this Section is to propose examples of perfectly transmitting chains, besides the exactly solved chain described by Eqs.~\eqref{e.MiKi}, and compare their dynamics with that of the uniform chain and two kinds of quasi-uniform chains described in Ref.~\cite{Vaia2018}: the first one with optimized $m_1$ and the second one with optimized $m_1$, $m_2$ and $K_1$. These optimized chains were shown to be able to rise the asymptotic transmission amplitude, which vanishes for the uniform chain, to $\alpha_\infty=0.847$ and to $\alpha_\infty=0.987$, respectively. Note that the uniform chain (with all $m_i=1$ and $K_i=1$) has the normal-mode frequencies
\begin{equation}
 \omega_n=2\sin\frac{k_n}2 ~,~~~ k_n=\frac{\pi(n{-}1)}{N},~~~~n=1,...,N ~,
\end{equation}
which are not commensurate, of course: the only low-$n$ modes are approximately spaced by $\omega=\pi/N$, while the spacing decreases for higher $n$. The parameters $k_n$ are quasi-wavevectors and lead to defining an analog of the group velocity~\cite{Vaia2018}, $\partial\omega/\partial{k}=\cos\frac{k}2$, which is almost one at low $k$: indeed, it turns out that the maximal transmission amplitude occurs after the pulse has been traveling along the chain for a time $t^*\gtrsim\pi/\omega$ that is slightly larger than $N$. The fact that low-$k$ modes are almost commensurate means that long-wavelength pulses travel more coherently: this corresponds to the vibrating-string limit, $N\to\infty$ with $\omega_k\to{k}$.
A travel time of order $N$ is also obtained when the only chain ends are modified in order to improve the transmission performance~\cite{Vaia2018}.

Table~\ref{t.N11n} reports data concerning three chains of 11 masses, i.e., the uniform and the optimized quasi-uniform chains. The first row gives the ``arrival time'' $t^*$ where $\alpha_{N}(t)$ attains its maximum value $\alpha=\alpha_{_N}(t^*)$: these quantities were calculated numerically. Besides the corresponding mirror-symmetric sequence of masses and springs, the table columns report the square-amplitudes of the normal modes, which are determined by the initial condition~\eqref{e.q0p0}, as well as the ``coherence factors'' at arrival,
\begin{equation}
 c_n=\cos[\pi(n{-}1){-}\omega_nt^*] ,~~~~n=1,...,N ~;
\end{equation}
the latter represent which fraction of the initial amplitude of the $n$th mode contributes to the overall transmitted amplitude $\alpha=\sum_nU_{n1}^2c_n$. It appears that higher-$n$ modes are less efficient (or less coherent), which explains the strategy used in the optimized chains: these perform better because the initial configuration gives larger weight to low-$n$ modes, which are reciprocally more coherent~\cite{Vaia2018}. This improvement can be appreciated by looking at the dynamical evolution shown in the first three panels of Fig.~\ref{f.N11}, where the instantaneous momenta of all masses are reported at equal time intervals between $t=0$ and $t^*$ (as said in Sec.~\ref{s.transmission}, one can equivalently think of the elongations of each mass). See the Supplemental Material~\cite{supplmat} for animations.

The same analysis applies to the longer chains, $N=41$, reported in Table~\ref{t.N41n} and in Fig.~\ref{f.N41}; there, the difference between the three cases is more evident. During the evolution the initial pulse appears to propagate along the chain with almost unit velocity, while at the arrival time an increasing amount of energy is transferred to the last mass, and not ``dispersed'' along the chain.

As discussed in Sec.~\ref{s.transmission}, all perfectly transmitting chains can be characterized by the sequence of odd coprime integers~\eqref{e.deltan} which identifies the spectrum~\eqref{e.om_kn}. In order to make a reasonable comparison with the above quasi-uniform chains, one can conveniently choose the parameter $\omega=\pi/(N{-}1)$, meaning that a pulse is expected to reach the opposite end at the transmission time $t^*=N{-}1$, equal to the chain length, hence with unit (average) velocity. Tables~\ref{t.N11p} and~\ref{t.N41p} report, besides the sequence~\eqref{e.deltan}, the corresponding values of masses and springs. In both tables, column (D) refers to the case of Eq.~\eqref{e.linear}, yielding the analytic recipe~\eqref{e.MiKi}, and the corresponding dynamics is shown in the fourth panels of Figs.~\ref{f.N11} and~\ref{f.N41}.
Columns (E) and (F) of the same tables report particular choices of the frequency sequence, among the infinite possible ones, which display less imbalanced (or ``more uniform'') chains, as quantified in the last row. The reported data were calculated numerically by the method described in Sec.~\ref{s.inverse}.

From the dynamics of these chains, Figs.~\ref{f.N11p} and~\ref{f.N41p}, it appears that the behavior is more complex, since a few of the normal modes are tuned with a frequency spacing $\delta>1$. As a matter of fact, if all $\delta_n$ were equal to $\delta$, they would not be coprime and one should set  $\omega\,\delta=\tilde\omega$, getting a shorter arrival time $\tilde{t}^*={t^*}/\delta$.
In the cases shown in  Figs.~\ref{f.N11} and~\ref{f.N41}, with $\delta=3$ (E) and $\delta=5$ (F), one indeed observes a faster propagation of the initial pulse, as the incipient behavior is to yield large transmission at the earlier time $\tilde{t}^*$; however, at this time not all modes are coherent yet: this is particularly appreciable in Fig.~\ref{f.N41p}, panel (F) at $t=t^*/5$; of course, perfect coherence among all modes will occur later at $t^*$.

\section{Conclusions}

Historically, the research aimed at obtaining efficient pulse transmission along one-dimensional mass-spring arrays mostly assumed translation-invariant (i.e., uniform) chains with nonelastic springs, as many nonlinear dynamic equations have been known to admit localized soliton-like solutions able to travel along the chain while preserving their shape.

At variance with this approach, in this paper uniformity is renounced instead of linearity, and it is shown how to characterize all mass-spring chains that yield perfect end-to-end pulse transmission, thus showing a dynamics analogous to that of a Newton's cradle.
In particular, all perfectly transmitting arrays of $N$ masses pairwise connected by $N\,{-}\,1$ springs are in one-to-one correspondence with the ordered sequences of coprime odd integers $(\delta_1,...,\,\delta_{N-1})$. For the simplest sequence $(1,\,1,\,...,1)$ there exists an analytic recipe~\cite{VaiaS2020}, while in the general case an efficient algorithm allows one to compute the mass-spring sequence numerically.

These ``magic'' chains can be used to build mechanical devices able, say, to efficiently transfer energy or momentum between the chain ends, whatever the chain length $N$. For instance, one can think of a desk toy that could replace the Newton cradle, like that shown in Fig.~\ref{f.cradle5}, the main difference being in the finite time required for a pulse to travel along the chain: in the ideal case where dissipation is neglected a pulse starting from one end can bounce back and forth indefinitely.

In the electrical circuit of Fig.~\ref{f.LCcircuit}, provided the capacitance-induction sequence is a ``magic'' one, a current pulse, generated by an external inductance coupled to $L_1$, would also bounce back and forth along the array.

One can imagine other applications, spanning from the macroscopic to the microscopic world~\cite{CahillEtAl2003,NorrisLB2013,RammPH2014}. The versatility of the model~\eqref{e.Horig} allows for many alternative implementations: basically, one could state that what is presented here constitutes a solution looking for a problem.
A suggestion may be that some biologically active polymer chains could be close to some ``magic'' mass-spring sequence, enhancing their ability to transfer energy.

\acknowledgments

The author thanks L.~Banchi, T.~J.~G.~Apollaro, A.~Cuccoli, and P.~Verrucchi for having driven his interest towards these topics.

\appendix
\section{Eqs.~(\ref{e.ex.aibi}) entail Eq.~(\ref{e.lambdaN})}
\label{a.proof}

Following Ref.~\cite{VaiaS2020}, the proof by induction on $N$ starts from $N\,{=}\,1$, where the statement is trivially true, as $\bm{A}_1=[0]_1$ has the eigenvalue $\lambda_1=0$. Assuming that the statement holds true for dimension $N$, one has to show that the validity follows for $N+1$, i.e., that $\bm{A}_{N+1}$ has the $N$ eigenvalues of $\bm{A}_N$ plus the eigenvalue $\lambda_{N+1}=2N^2$.

The entries of $\bm{A}_{N+1}$ are
\begin{equation}
\begin{aligned}
 a_i&=N+4(i{-}1)(N{+}1{-}i) ~, &&i=1,\,...,\,N{+}1~,
\\
 b_i^2&=i\,(2i{-}1)\,(N{+}1{-}i)\,(2N{+}1{-}2i) ~,&&i=1,\,...,\,N~.
\label{e.ex.aibi1}
\end{aligned}
\end{equation}
It is simple algebra to verify that the tridiagonal matrix $2N^2-\bm{A}_{N+1}$, with diagonal elements
\begin{equation}
 2N^2-a_i=N(N{-}1)+(N{+}2{-}2i)^2 ~,
\end{equation}
factorizes as $2N^2-\bm{A}_{N+1}=\bm{H}\bm{H}^{\rm T}$. The matrix
\begin{equation}
\bm{H} = 
\begin{bmatrix}
	 h_1   &  0     &\cdots   &  0      \\
	 r_1   &  h_2   &         & \vdots  \\
	\vdots &        &~\ddots  &  0      \\
	 0     &\cdots  &  r_N    &  h_{N+1}
\end{bmatrix}_{N+1}
\label{e.HN1}
\end{equation}
is lower bidiagonal and has positive elements given by
\begin{equation}
\begin{aligned}
 h_i^2&=(N{+}1{-}i)(2N{+}1{-}2i) ~, &&i=1,\,...,\,N{+}1~,
\\
 r_i^2&=i\,(2i{-}1) ~,&&i=1,\,...,\,N~.
\label{e.ex.hiri}
\end{aligned}
\end{equation}
The matrix $\bm{A}_{N+1}=2N^2-\bm{H}\bm{H}^{\rm T}$, has the same spectrum of the matrix
\begin{equation}
 2N^2-\bm{H}^{\rm T}\bm{H} = 
\begin{bmatrix}
	    &          &   &  0      \\
	    & ~\bm{A}_N &   &  \vdots \\
	    &          &   &  0      \\
	 0  & \cdots    & 0 &  2N^2
\end{bmatrix}_{N+1}~;
\end{equation}
hence, its eigenvalues are those of $\bm{A}_N$ plus the $(N+1)$th eigenvalue $2N^2$.

\newpage \phantom{.} \newpage

\onecolumngrid

\setlength\tabcolsep{3mm}

\begin{table}
\caption{Comparison of different mass-spring chains with $N=11$. The first column (A) refers to the uniform chain, the next two columns (B, C) to the optimized chains studied in Ref.~\cite{Vaia2018}; $t^*$ is the ``arrival time''  when the transmission amplitude $\alpha=\alpha_{_N}(t^*)$ is maximal; $c_n=\cos[\pi(n{-}1){-}\omega_nt^*]$ is the coherence factor and $U_{n1}^2$ is the weight of each mode. Transmission is increasingly efficient from (A) to (C) because heavier coherence factors are closer to unity.
}
\label{t.N11n}
\medskip
\begin{tabular}{|r|llll|llll|llll|}
 \hline \rule{0pt}{10pt}
		& \multicolumn{4}{c|}{(A) uniform}
        & \multicolumn{4}{c|}{(B) optimal $m_1$}
        & \multicolumn{4}{c|}{(C) optimal  $m_1,\,m_2,\,K_1$} 
\\ 
		& \multicolumn{4}{c|}{$t^*=11.917$~~~~$\alpha=0.787$}
        & \multicolumn{4}{c|}{$t^*=13.039$~~~~$\alpha=0.972$}
        & \multicolumn{4}{c|}{$t^*=13.351$~~~~$\alpha=0.989$} 
\\ \hline \rule[-3pt]{0pt}{13pt}
  $i,n$ & $m_i$  & $K_i$  & $c_n$~~    & $U_{n1}^2$
        & $m_i$~ & $K_i$  & $c_n$~~    & $U_{n1}^2$
        & $m_i$~ & $K_i$~ & $c_n$~~    & $U_{n1}^2$ \\
 \hline\rule{0pt}{9pt}
 1 & 1 & 1 & 1.0000 & 0.0909 & 2.4121 &  1 & 1.0000 & 0.1745 & 2.1259 & 0.7212 & 1.0000 & 0.1639 \\
 2 & 1 & 1 & 0.9688 & 0.1781 & 1      &  1 & 0.9902 & 0.3046 & 0.8606 & 1      & 0.9988 & 0.3084 \\
 3 & 1 & 1 & 0.9083 & 0.1674 & 1      &  1 & 0.9943 & 0.2121 & 1      & 1      & 0.9992 & 0.2418 \\
 4 & 1 & 1 & 0.8887 & 0.1504 & 1      &  1 & 0.9969 & 0.1311 & 1      & 1      & 0.9945 & 0.1467 \\
 5 & 1 & 1 & 0.9495 & 0.1287 & 1      &  1 & 0.9646 & 0.0781 & 1      & 1      & 0.9992 & 0.0723 \\
 6 & 1 & 1 & 0.9950 & 0.1039 & 1      &  1 & 0.9566 & 0.0462 & 1      & 1      & 0.9633 & 0.0338 \\
 7 & 1 & 1 & 0.6697 & 0.0780 & 1      &  1 & 0.9994 & 0.0269 & 1      & 1      & 0.9515 & 0.0167 \\
 8 & 1 & 1 & 0.3615 & 0.0531 & 1      &  1 & 0.8130 & 0.0150 & 1      & 1      & 0.9968 & 0.0089 \\
 9 & 1 & 1 & 0.9520 & 0.0314 & 1      &  1 & 0.1771 & 0.0077 & 1      & 1      & 0.4874 & 0.0048 \\
10 & 1 & 1 & 0.6392 & 0.0144 & 1      &  1 & 0.9645 & 0.0032 & 0.8606 & 0.7212 & 0.9048 & 0.0022 \\
11 & 1 &   & 0.0295 & 0.0037 & 2.4121 &    & 0.8030 & 0.0008 & 2.1259 &        & 0.2713 & 0.0006 \\
\hline
	\end{tabular}
\end{table}

\begin{table}
\caption{Comparison of different perfectly transmitting mass-spring chains with $N=11$. The first column (D) refers to the perfect chain~\cite{VaiaS2020} with the spectrum~\eqref{e.linear}, columns (E) and (F) are sample cases with different frequency spacings~$\delta_n$, Eq.\eqref{e.deltan}. The last row reports the ratio between largest and smallest values, quantifying the non-uniformity of the chain.
}
\label{t.N11p}
\medskip
\begin{tabular}{|r|rrrr|rrrr|rrrr|}
 \hline \rule{0pt}{10pt}
		& \multicolumn{4}{c|}{(D)}
        & \multicolumn{4}{c|}{(E)}
        & \multicolumn{4}{c|}{(F)} 
\\ \hline \rule[-3pt]{0pt}{13pt}
  $i,n$ & $\delta_n$ & $k_n$ & $m_i$~ & $K_i$~
        & $\delta_n$ & $k_n$ & $m_i$~ & $K_i$~
        & $\delta_n$ & $k_n$ & $m_i$~ & $K_i$~  \\
 \hline \rule{0pt}{9pt}
 1 & 1 & 0 & 1.0000 & 0.4935 & 3 & 0 & 1.0000 & 13.068 & 5 & 0 & 1.0000 & 39.020 \\
 2 & 1 & 1 & 0.5263 & 0.7013 & 3 & 3 & 1.0983 & 11.014 & 5 & 5 & 0.9636 & 28.190 \\
 3 & 1 & 2 & 0.4180 & 0.8250 & 3 & 6 & 0.8129 & 9.925 & 5 & 10 & 0.8206 & 27.888 \\
 4 & 1 & 3 & 0.3715 & 0.8983 & 3 & 9 & 1.0745 & 11.539 & 5 & 15 & 0.9779 & 27.903 \\
 5 & 1 & 4 & 0.3501 & 0.9329 & 3 & 12 & 0.7550 & 10.746 & 5 & 20 & 0.8471 & 29.662 \\
 6 & 1 & 5 & 0.3437 & 0.9329 & 1 & 15 & 1.0727 & 10.746 & 3 & 25 & 0.8905 & 29.662 \\
 7 & 1 & 6 & 0.3501 & 0.8983 & 3 & 16 & 0.7550 & 11.539 & 3 & 28 & 0.8471 & 27.903 \\
 8 & 1 & 7 & 0.3715 & 0.8250 & 1 & 19 & 1.0745 & 9.925 & 3 & 31 & 0.9779 & 27.888 \\
 9 & 1 & 8 & 0.4180 & 0.7013 & 1 & 20 & 0.8129 & 11.014 & 1 & 34 & 0.8206 & 28.190 \\
10 & 1 & 9 & 0.5263 & 0.4935 & 1 & 21 & 1.0983 & 13.068 & 1 & 35 & 0.9636 & 39.020 \\
11 &  & 10 & 1.0000 &        &   & 22 & 1.0000 &        &   & 36 & 1.0000 &  \\
\hline \rule[-6pt]{0pt}{18pt}
 $\frac{\rm{max}}{\rm{min}}$ & & & 2.91 & 1.89 & & & 1.45 & 1.32 & & & 1.22 & 1.40 \\
\hline
	\end{tabular}
\end{table}

\begin{table}
\caption{Comparison of different mass-spring chains with $N=41$. The first column (A) refers to the uniform chain, the next two columns (B, C) to the optimized chains studied in Ref.~\cite{Vaia2018}; $t^*$ is the ``arrival time''  when the transmission amplitude $\alpha=\alpha_{_N}(t^*)$ is maximal; $c_n=\cos[\pi(n{-}1){-}\omega_nt^*]$ is the coherence factor and $U_{n1}^2$ is the weight of each mode. Transmission is increasingly efficient from (A) to (C) because heavier coherence factors are closer to unity.
}
\label{t.N41n}
\medskip
\begin{tabular}{|r|llll|llll|llll|}
 \hline \rule{0pt}{10pt}
		& \multicolumn{4}{c|}{(A) uniform}
        & \multicolumn{4}{c|}{(B) optimal $m_1$}
        & \multicolumn{4}{c|}{(C) optimal  $m_1,\,m_2,\,K_1$} 
\\ 
		& \multicolumn{4}{c|}{$t^*=42.620$~~~~$\alpha=0.5681$}
        & \multicolumn{4}{c|}{$t^*=44.787$~~~~$\alpha=0.9377$}
        & \multicolumn{4}{c|}{$t^*=45.702$~~~~$\alpha=0.9844$} 
\\ \hline \rule[-3pt]{0pt}{13pt}
  $i,n$ & $m_i$  & $K_i$  & $c_n$~~    & $U_{n1}^2$
        & $m_i$~ & $K_i$  & $c_n$~~    & $U_{n1}^2$
        & $m_i$~ & $K_i$~ & $c_n$~~    & $U_{n1}^2$ \\
 \hline\rule{0pt}{9pt}
 1 & 1 & 1 & 1.0000 & 0.0244 & 3.8133 & 1 & 1.0000 & 0.0818 & 3.2017 & 0.5255 & 1.0000 & 0.0713\\
 2 & 1 & 1 & 0.9924 & 0.0487 & 1 & 1 & 0.9931 & 0.1568 & 0.7627 & 1 & 0.9988 & 0.1420\\
 3 & 1 & 1 & 0.9709 & 0.0485 & 1 & 1 & 0.9798 & 0.1392 & 1 & 1 & 0.9974 & 0.1396\\
 4 & 1 & 1 & 0.9391 & 0.0481 & 1 & 1 & 0.9734 & 0.1166 & 1 & 1 & 0.9986 & 0.1330\\
 5 & 1 & 1 & 0.9024 & 0.0476 & 1 & 1 & 0.9790 & 0.0943 & 1 & 1 & 1.0000 & 0.1200\\
 6 & 1 & 1 & 0.8673 & 0.0470 & 1 & 1 & 0.9911 & 0.0751 & 1 & 1 & 0.9970 & 0.1010\\
 7 & 1 & 1 & 0.8404 & 0.0463 & 1 & 1 & 0.9995 & 0.0596 & 1 & 1 & 0.9914 & 0.0791\\
 8 & 1 & 1 & 0.8276 & 0.0454 & 1 & 1 & 0.9965 & 0.0475 & 1 & 1 & 0.9901 & 0.0584\\
 9 & 1 & 1 & 0.8331 & 0.0443 & 1 & 1 & 0.9803 & 0.0382 & 1 & 1 & 0.9951 & 0.0416\\
10 & 1 & 1 & 0.8585 & 0.0432 & 1 & 1 & 0.9549 & 0.0310 & 1 & 1 & 0.9999 & 0.0292\\
11 & 1 & 1 & 0.9011 & 0.0420 & 1 & 1 & 0.9285 & 0.0254 & 1 & 1 & 0.9955 & 0.0205\\
12 & 1 & 1 & 0.9518 & 0.0406 & 1 & 1 & 0.9104 & 0.0210 & 1 & 1 & 0.9779 & 0.0146\\
13 & 1 & 1 & 0.9924 & 0.0392 & 1 & 1 & 0.9082 & 0.0175 & 1 & 1 & 0.9511 & 0.0106\\
14 & 1 & 1 & 0.9929 & 0.0377 & 1 & 1 & 0.9253 & 0.0147 & 1 & 1 & 0.9250 & 0.0078\\
15 & 1 & 1 & 0.9120 & 0.0360 & 1 & 1 & 0.9579 & 0.0124 & 1 & 1 & 0.9111 & 0.0059\\
16 & 1 & 1 & 0.7039 & 0.0344 & 1 & 1 & 0.9912 & 0.0105 & 1 & 1 & 0.9179 & 0.0045\\
17 & 1 & 1 & 0.3389 & 0.0326 & 1 & 1 & 0.9957 & 0.0089 & 1 & 1 & 0.9463 & 0.0035\\
18 & 1 & 1 & 0.1606 & 0.0309 & 1 & 1 & 0.9247 & 0.0077 & 1 & 1 & 0.9838 & 0.0028\\
19 & 1 & 1 & 0.6774 & 0.0290 & 1 & 1 & 0.7211 & 0.0066 & 1 & 1 & 0.9989 & 0.0023\\
20 & 1 & 1 & 0.9866 & 0.0272 & 1 & 1 & 0.3416 & 0.0056 & 1 & 1 & 0.9363 & 0.0019\\
21 & 1 & 1 & 0.8355 & 0.0253 & 1 & 1 & 0.1937 & 0.0048 & 1 & 1 & 0.7231 & 0.0016\\
22 & 1 & 1 & 0.1560 & 0.0235 & 1 & 1 & 0.7377 & 0.0042 & 1 & 1 & 0.3024 & 0.0013\\
23 & 1 & 1 & 0.6933 & 0.0216 & 1 & 1 & 0.9999 & 0.0036 & 1 & 1 & 0.2922 & 0.0011\\
24 & 1 & 1 & 0.9778 & 0.0198 & 1 & 1 & 0.6837 & 0.0031 & 1 & 1 & 0.8459 & 0.0009\\
25 & 1 & 1 & 0.2445 & 0.0179 & 1 & 1 & 0.1784 & 0.0026 & 1 & 1 & 0.9659 & 0.0008\\
26 & 1 & 1 & 0.8221 & 0.0162 & 1 & 1 & 0.9385 & 0.0022 & 1 & 1 & 0.3595 & 0.0007\\
27 & 1 & 1 & 0.7592 & 0.0144 & 1 & 1 & 0.6993 & 0.0019 & 1 & 1 & 0.6402 & 0.0006\\
28 & 1 & 1 & 0.5289 & 0.0127 & 1 & 1 & 0.4688 & 0.0016 & 1 & 1 & 0.9520 & 0.0005\\
29 & 1 & 1 & 0.8711 & 0.0111 & 1 & 1 & 0.9566 & 0.0013 & 1 & 1 & 0.0558 & 0.0005\\
30 & 1 & 1 & 0.5475 & 0.0096 & 1 & 1 & 0.2207 & 0.0011 & 1 & 1 & 0.9977 & 0.0004\\
31 & 1 & 1 & 0.7266 & 0.0082 & 1 & 1 & 0.9635 & 0.0009 & 1 & 1 & 0.0585 & 0.0004\\
32 & 1 & 1 & 0.8673 & 0.0068 & 1 & 1 & 0.4295 & 0.0007 & 1 & 1 & 0.9962 & 0.0003\\
33 & 1 & 1 & 0.1124 & 0.0056 & 1 & 1 & 0.7364 & 0.0006 & 1 & 1 & 0.3627 & 0.0003\\
34 & 1 & 1 & 0.9158 & 0.0044 & 1 & 1 & 0.9210 & 0.0005 & 1 & 1 & 0.7009 & 0.0002\\
35 & 1 & 1 & 0.8717 & 0.0034 & 1 & 1 & 0.1569 & 0.0003 & 1 & 1 & 0.9772 & 0.0002\\
36 & 1 & 1 & 0.2512 & 0.0025 & 1 & 1 & 0.6621 & 0.0003 & 1 & 1 & 0.4694 & 0.0002\\
37 & 1 & 1 & 0.4161 & 0.0018 & 1 & 1 & 0.9951 & 0.0002 & 1 & 1 & 0.2383 & 0.0001\\
38 & 1 & 1 & 0.8357 & 0.0011 & 1 & 1 & 0.8706 & 0.0001 & 1 & 1 & 0.7394 & 0.0001\\
39 & 1 & 1 & 0.9895 & 0.0006 & 1 & 1 & 0.5464 & 0.0001 & 1 & 1 & 0.9604 & 0.0001\\
40 & 1 & 1 & 0.9860 & 0.0003 & 1 & 1 & 0.2337 & 0.0000 & 0.7627 & 0.5255 & 0.9992 & 0.0000\\
41 & 1 &   & 0.9377 & 0.0001 & 3.8133 &  & 0.0304 & 0.0000 & 3.2017 &  & 0.9728 & 0.0000\\
\hline
	\end{tabular}
\end{table}

\begin{table}
\caption{Comparison of different perfectly transmitting mass-spring chains with $N=41$. The first column (D) refers to the perfect chain~\cite{VaiaS2020} with the spectrum~\eqref{e.linear}, columns (E) and (F) are sample cases with different frequency spacings~$\delta_n$, Eq.\eqref{e.deltan}. The last row reports the ratio between largest and smallest values, quantifying the non-uniformity of the chain.
}
\label{t.N41p}
\medskip
\begin{tabular}{|r|rrrr|rrrr|rrrr|}
 \hline \rule{0pt}{10pt}
		& \multicolumn{4}{c|}{(D)}
        & \multicolumn{4}{c|}{(E)}
        & \multicolumn{4}{c|}{(F)} 
\\ \hline \rule[-3pt]{0pt}{13pt}
  $i,n$ & $\delta_n$ & $k_n$ & $m_i$~ & $K_i$~
        & $\delta_n$ & $k_n$ & $m_i$~ & $K_i$~
        & $\delta_n$ & $k_n$ & $m_i$~ & $K_i$~  \\
 \hline \rule{0pt}{9pt}
 1 & 1 &  0 & 1.0000 & 0.1234 & 3 &  0 & 1.0000 & 6.413 & 5 &   0 & 1.0000 & 22.240 \\
 2 & 1 &  1 & 0.5063 & 0.1827 & 3 &  3 & 1.2055 & 2.677 & 5 &   5 & 1.0582 & 14.851 \\
 3 & 1 &  2 & 0.3847 & 0.2254 & 3 &  6 & 1.0458 & 3.775 & 5 &  10 & 1.0080 & 20.927 \\
 4 & 1 &  3 & 0.3248 & 0.2595 & 3 &  9 & 0.8147 & 4.351 & 5 &  15 & 0.8424 & 15.130 \\
 5 & 1 &  4 & 0.2881 & 0.2879 & 3 & 12 & 0.6823 & 6.066 & 5 &  20 & 0.7170 & 20.767 \\
 6 & 1 &  5 & 0.2630 & 0.3123 & 3 & 15 & 0.4889 & 7.352 & 5 &  25 & 0.8666 & 23.079 \\
 7 & 1 &  6 & 0.2445 & 0.3334 & 3 & 18 & 0.4585 & 6.310 & 5 &  30 & 0.6978 & 21.220 \\
 8 & 1 &  7 & 0.2305 & 0.3519 & 3 & 21 & 0.5433 & 7.094 & 5 &  35 & 0.6676 & 20.974 \\
 9 & 1 &  8 & 0.2194 & 0.3681 & 3 & 24 & 0.4688 & 7.156 & 5 &  40 & 0.6944 & 21.126 \\
10 & 1 &  9 & 0.2105 & 0.3824 & 3 & 27 & 0.4619 & 6.484 & 5 &  45 & 0.7305 & 23.319 \\
11 & 1 & 10 & 0.2032 & 0.3949 & 3 & 30 & 0.5269 & 6.960 & 3 &  50 & 0.7411 & 22.081 \\
12 & 1 & 11 & 0.1973 & 0.4059 & 3 & 33 & 0.4952 & 7.083 & 5 &  53 & 0.6724 & 20.295 \\
13 & 1 & 12 & 0.1924 & 0.4154 & 3 & 36 & 0.4399 & 6.955 & 5 &  58 & 0.6963 & 21.596 \\
14 & 1 & 13 & 0.1884 & 0.4235 & 3 & 39 & 0.4906 & 7.097 & 5 &  63 & 0.7353 & 22.397 \\
15 & 1 & 14 & 0.1851 & 0.4303 & 3 & 42 & 0.5095 & 6.891 & 5 &  68 & 0.7169 & 22.708 \\
16 & 1 & 15 & 0.1824 & 0.4360 & 1 & 45 & 0.4543 & 6.973 & 5 &  73 & 0.6856 & 21.483 \\
17 & 1 & 16 & 0.1803 & 0.4404 & 1 & 46 & 0.4654 & 7.369 & 5 &  78 & 0.6667 & 21.049 \\
18 & 1 & 17 & 0.1787 & 0.4437 & 1 & 47 & 0.4993 & 7.038 & 3 &  83 & 0.7254 & 22.458 \\
19 & 1 & 18 & 0.1776 & 0.4459 & 3 & 48 & 0.4667 & 6.760 & 3 &  86 & 0.7442 & 22.036 \\
20 & 1 & 19 & 0.1770 & 0.4470 & 3 & 51 & 0.4656 & 7.330 & 3 &  89 & 0.6879 & 21.780 \\
21 & 1 & 20 & 0.1767 & 0.4470 & 3 & 54 & 0.4898 & 7.330 & 3 &  92 & 0.6675 & 21.780 \\
22 & 1 & 21 & 0.1770 & 0.4459 & 3 & 57 & 0.4656 & 6.760 & 3 &  95 & 0.6879 & 22.036 \\
23 & 1 & 22 & 0.1776 & 0.4437 & 3 & 60 & 0.4667 & 7.038 & 3 &  98 & 0.7442 & 22.458 \\
24 & 1 & 23 & 0.1787 & 0.4404 & 3 & 63 & 0.4993 & 7.369 & 3 & 101 & 0.7254 & 21.049 \\
25 & 1 & 24 & 0.1803 & 0.4360 & 3 & 66 & 0.4654 & 6.973 & 3 & 104 & 0.6667 & 21.483 \\
26 & 1 & 25 & 0.1824 & 0.4303 & 3 & 69 & 0.4543 & 6.891 & 3 & 107 & 0.6856 & 22.708 \\
27 & 1 & 26 & 0.1851 & 0.4235 & 3 & 72 & 0.5095 & 7.097 & 3 & 110 & 0.7169 & 22.397 \\
28 & 1 & 27 & 0.1884 & 0.4154 & 3 & 75 & 0.4906 & 6.955 & 3 & 113 & 0.7353 & 21.596 \\
29 & 1 & 28 & 0.1924 & 0.4059 & 3 & 78 & 0.4399 & 7.083 & 3 & 116 & 0.6963 & 20.295 \\
30 & 1 & 29 & 0.1973 & 0.3949 & 3 & 81 & 0.4952 & 6.960 & 3 & 119 & 0.6724 & 22.081 \\
31 & 1 & 30 & 0.2032 & 0.3824 & 3 & 84 & 0.5269 & 6.484 & 3 & 122 & 0.7411 & 23.319 \\
32 & 1 & 31 & 0.2105 & 0.3681 & 1 & 87 & 0.4619 & 7.156 & 3 & 125 & 0.7305 & 21.126 \\
33 & 1 & 32 & 0.2194 & 0.3519 & 1 & 88 & 0.4688 & 7.094 & 3 & 128 & 0.6944 & 20.974 \\
34 & 1 & 33 & 0.2305 & 0.3334 & 3 & 89 & 0.5433 & 6.310 & 3 & 131 & 0.6676 & 21.220 \\
35 & 1 & 34 & 0.2445 & 0.3123 & 1 & 92 & 0.4585 & 7.352 & 1 & 134 & 0.6978 & 23.079 \\
36 & 1 & 35 & 0.2630 & 0.2879 & 1 & 93 & 0.4889 & 6.066 & 3 & 135 & 0.8666 & 20.767 \\
37 & 1 & 36 & 0.2881 & 0.2595 & 1 & 94 & 0.6823 & 4.351 & 1 & 138 & 0.7170 & 15.130 \\
38 & 1 & 37 & 0.3248 & 0.2254 & 1 & 95 & 0.8147 & 3.775 & 1 & 139 & 0.8424 & 20.927 \\
39 & 1 & 38 & 0.3847 & 0.1827 & 1 & 96 & 1.0458 & 2.677 & 1 & 140 & 1.0080 & 14.851 \\
40 & 1 & 39 & 0.5063 & 0.1234 & 1 & 97 & 1.2055 & 6.413 & 1 & 141 & 1.0582 & 22.240 \\
41 &  & 40 & 1.0000 &  & & 98 & 1.0000 &  &  & 142 & 1.0000 &  \\
\hline \rule[-6pt]{0pt}{18pt}
 $\frac{\rm{max}}{\rm{min}}$ & & & 5.66 & 3.62 & & & 2.74 & 2.75 & & & 1.59 & 1.57 \\
\hline
	\end{tabular}
\end{table}

\begin{figure}
\includegraphics[width=0.4\textwidth]{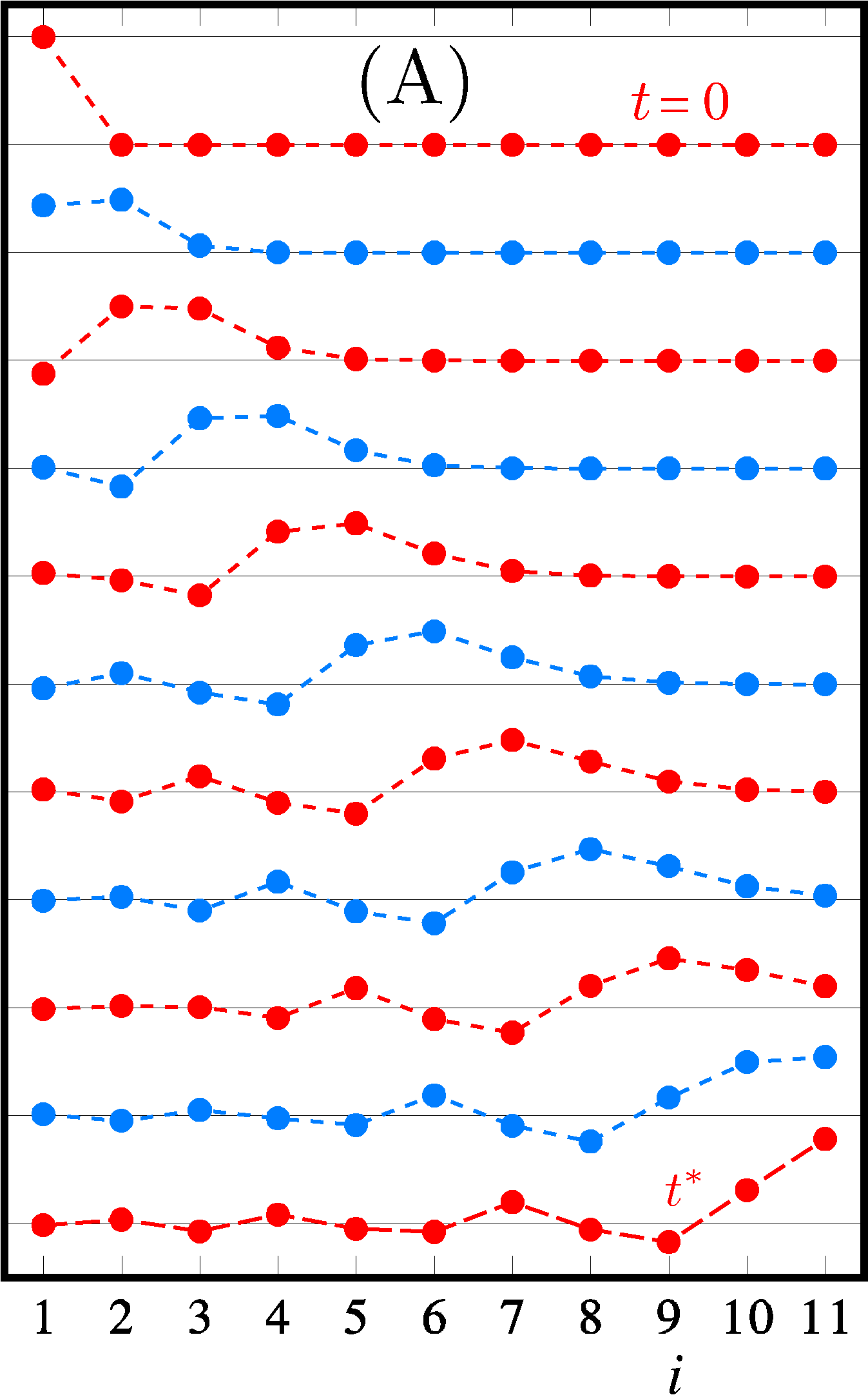}~~~~~
\includegraphics[width=0.4\textwidth]{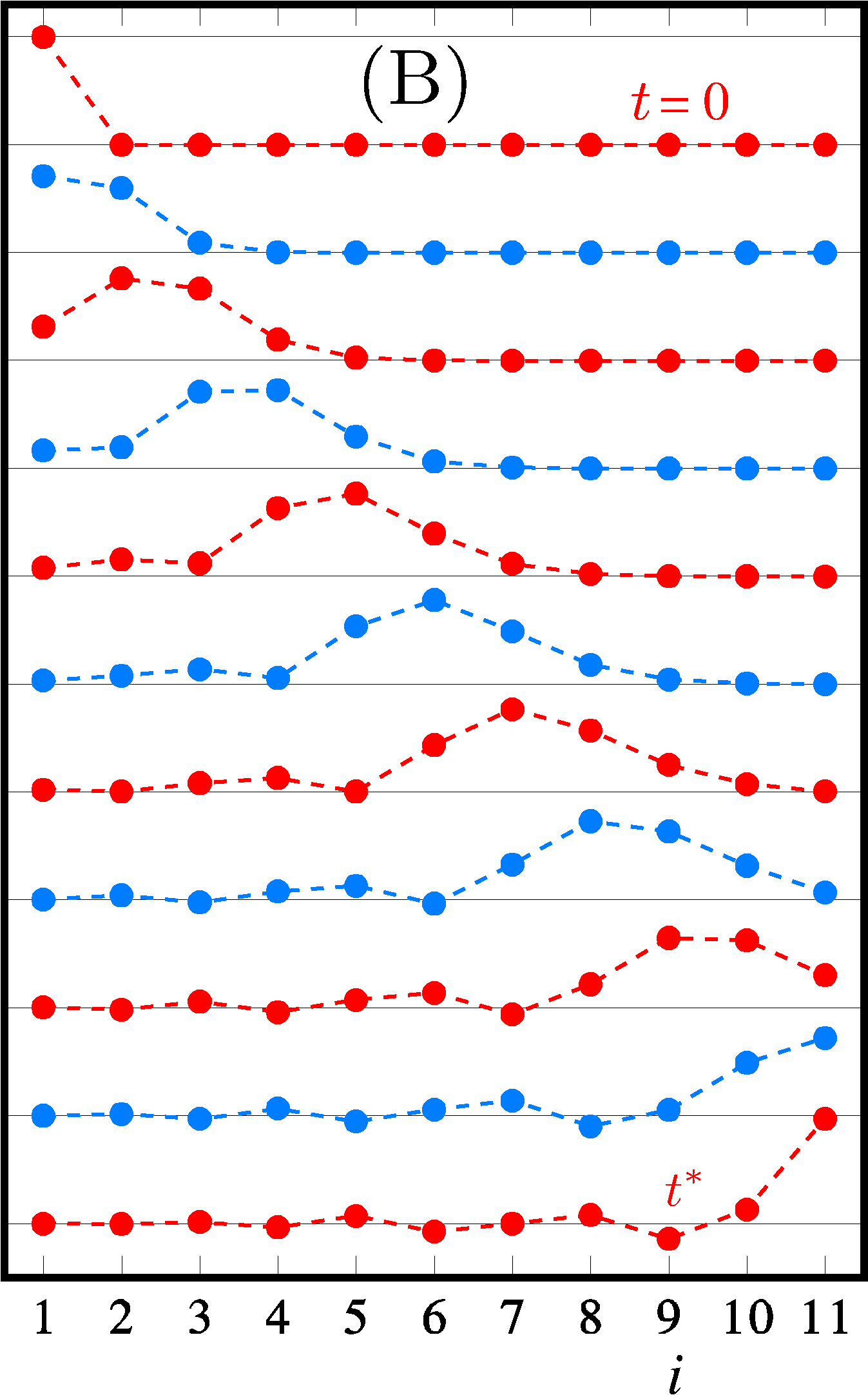}\\[3mm]
\includegraphics[width=0.4\textwidth]{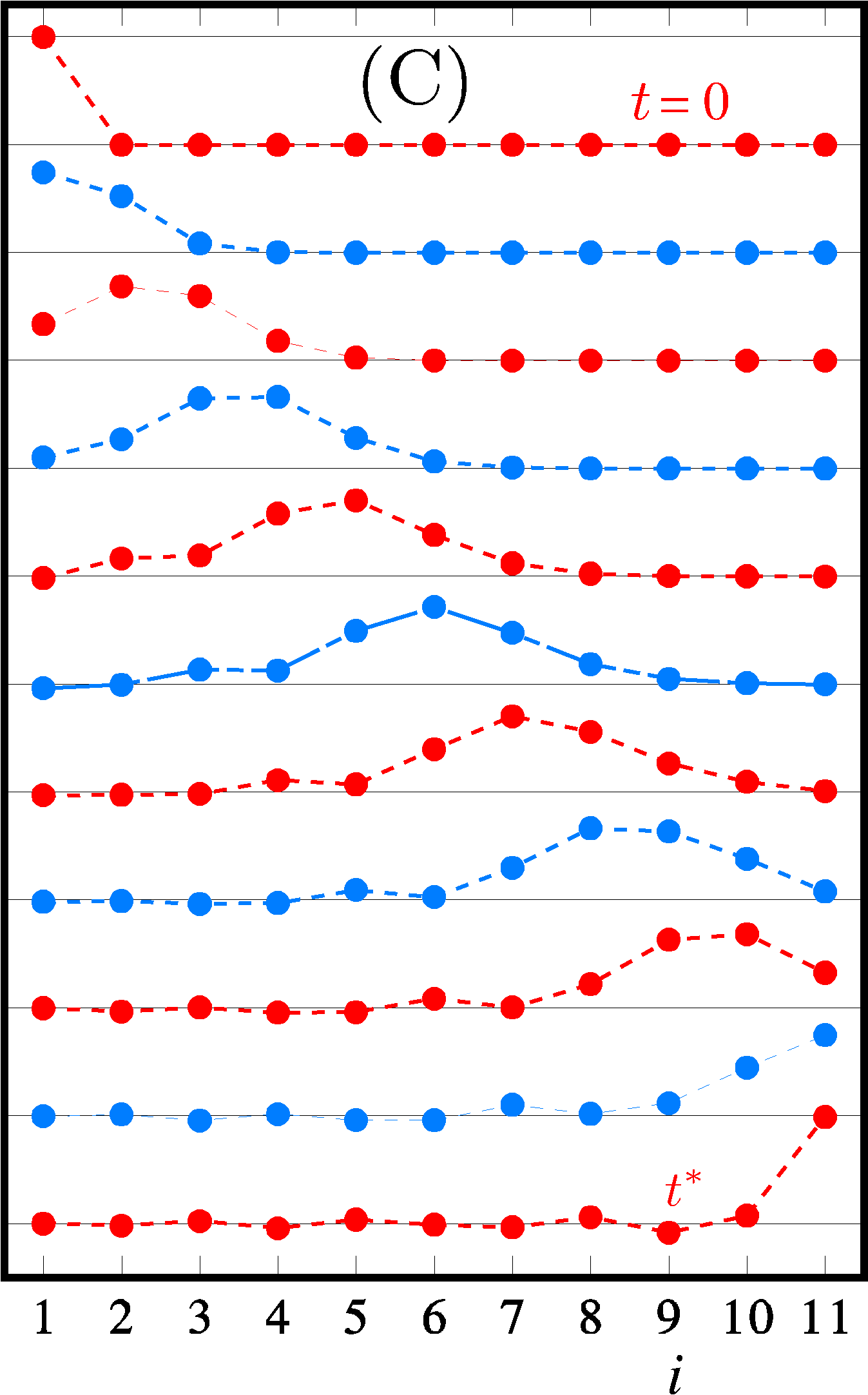}~~~~~
\includegraphics[width=0.4\textwidth]{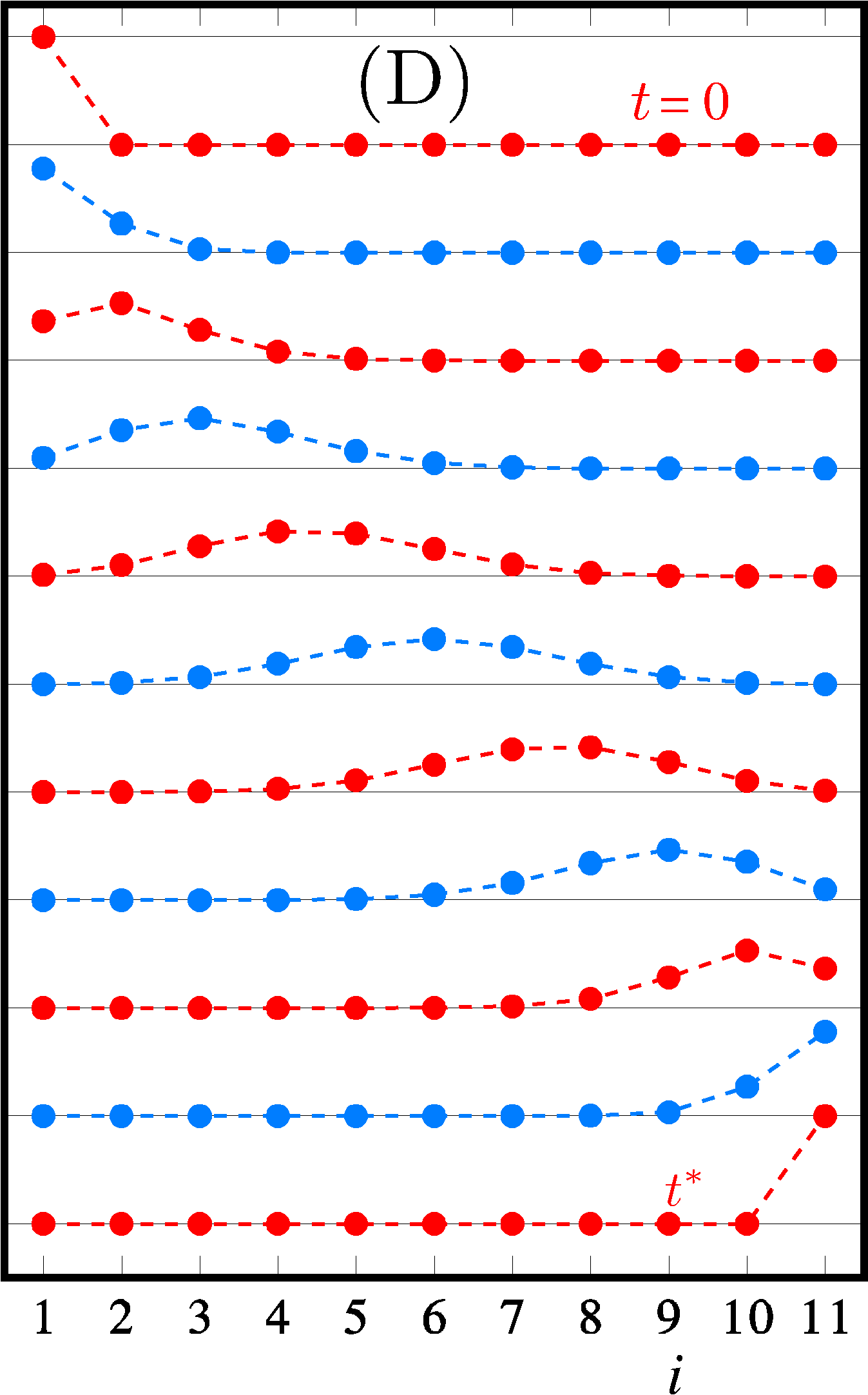}
\caption{$N=11$ chain, snapshots of the dynamics at equal time intervals of $t^*/10$ between $0$ and $t^*$. The ordinate represents the momenta $p_i(t)$ of each mass as they evolve starting from the configuration~\eqref{e.q0p0}.
The first three panels correspond to the columns of Table~\ref{t.N11n}, namely the uniform chain (A), the quasi-uniform chains with optimal extremal mass $m_1$ (B) and with 2 optimized extremal masses, $m_1,~m_2$, and their spring $K_1$ (C); the last panel is the perfect chain (D) reported in Table~\ref{t.N11p}. Animations are available as Supplemental Material~\cite{supplmat}. }
\label{f.N11}
\end{figure}

\begin{figure}
\includegraphics[width=0.38\textwidth]{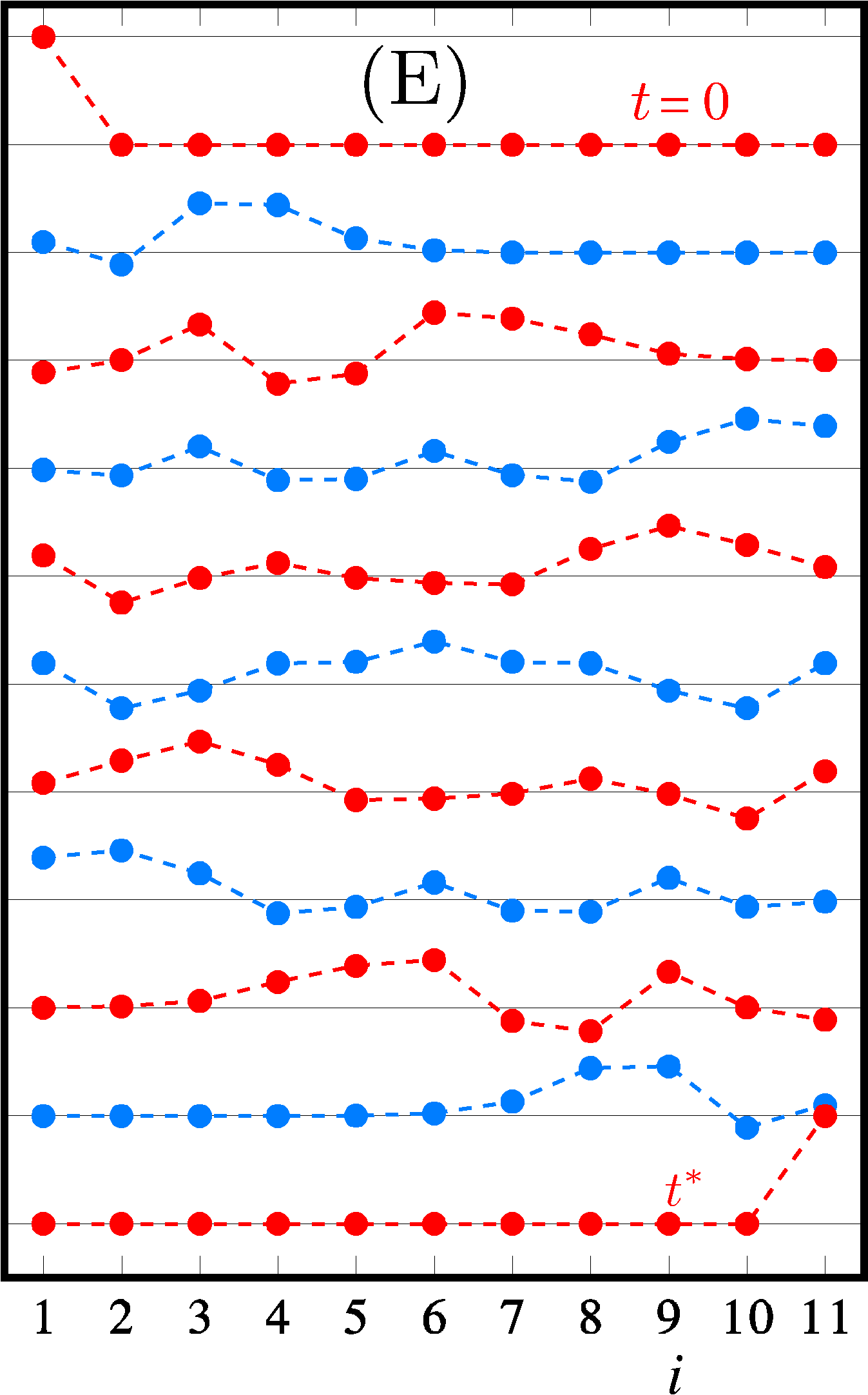}~~~~~
\includegraphics[width=0.38\textwidth]{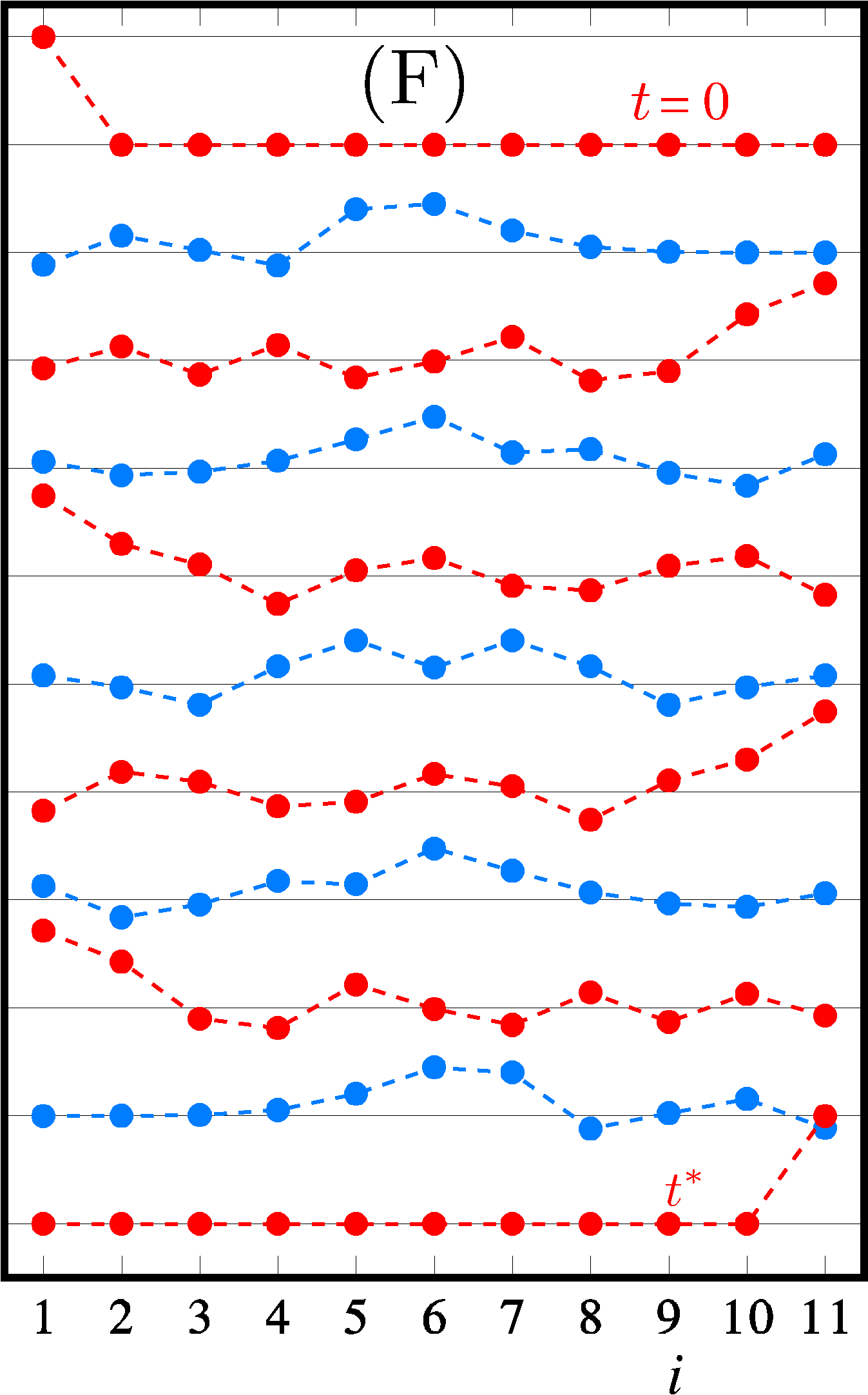}
\caption{Snapshots of the dynamics at equal time intervals of $t^*/10$ between $0$ and $t^*=10$ for the two alternative $N=11$ perfectly transmitting chains (E) and (F) described in Table~\ref{t.N11p}.  Animations are available as Supplemental Material~\cite{supplmat}. }
\label{f.N11p}
\end{figure}

\begin{figure}
\includegraphics[width=0.4\textwidth]{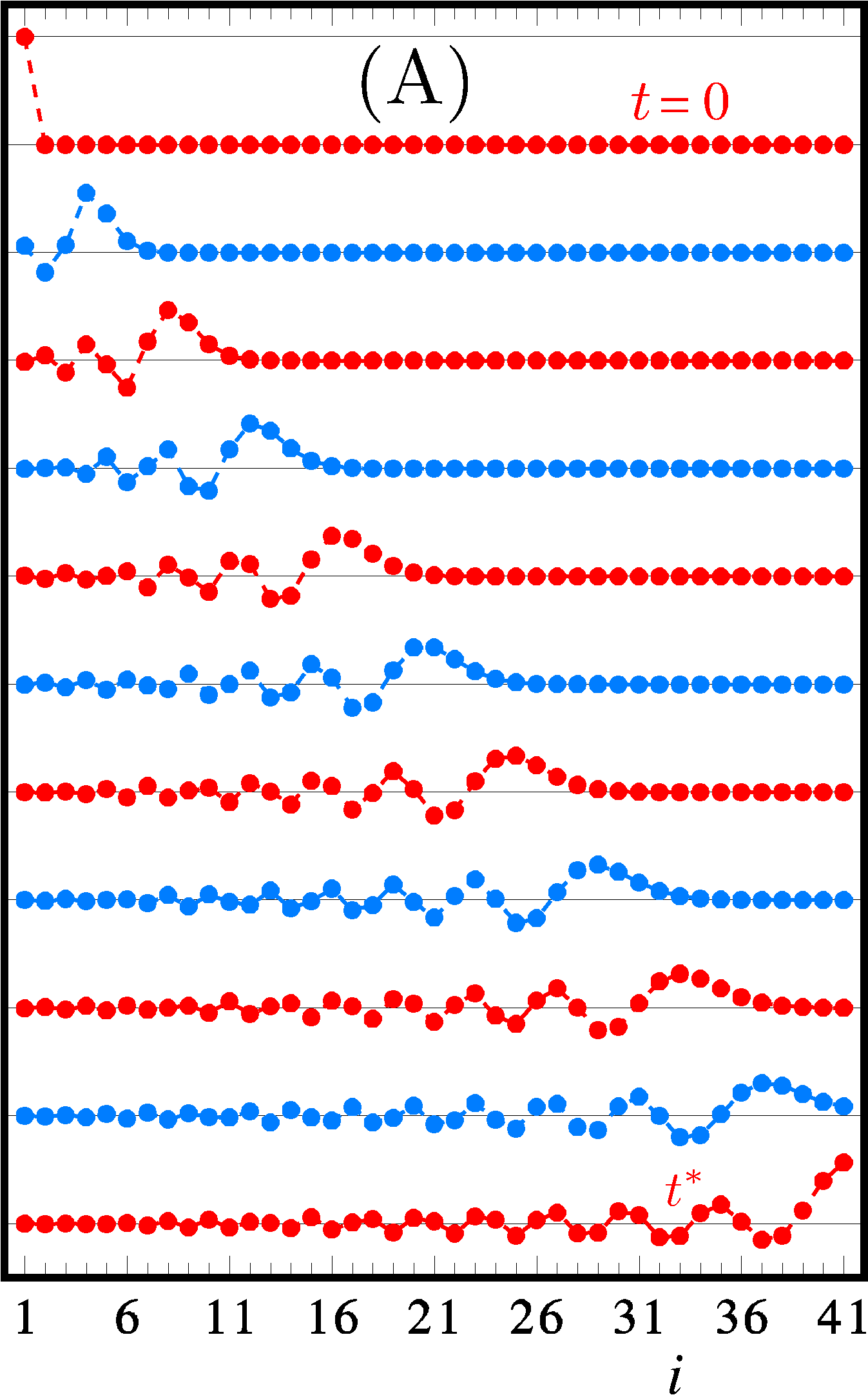}~~~~~
\includegraphics[width=0.4\textwidth]{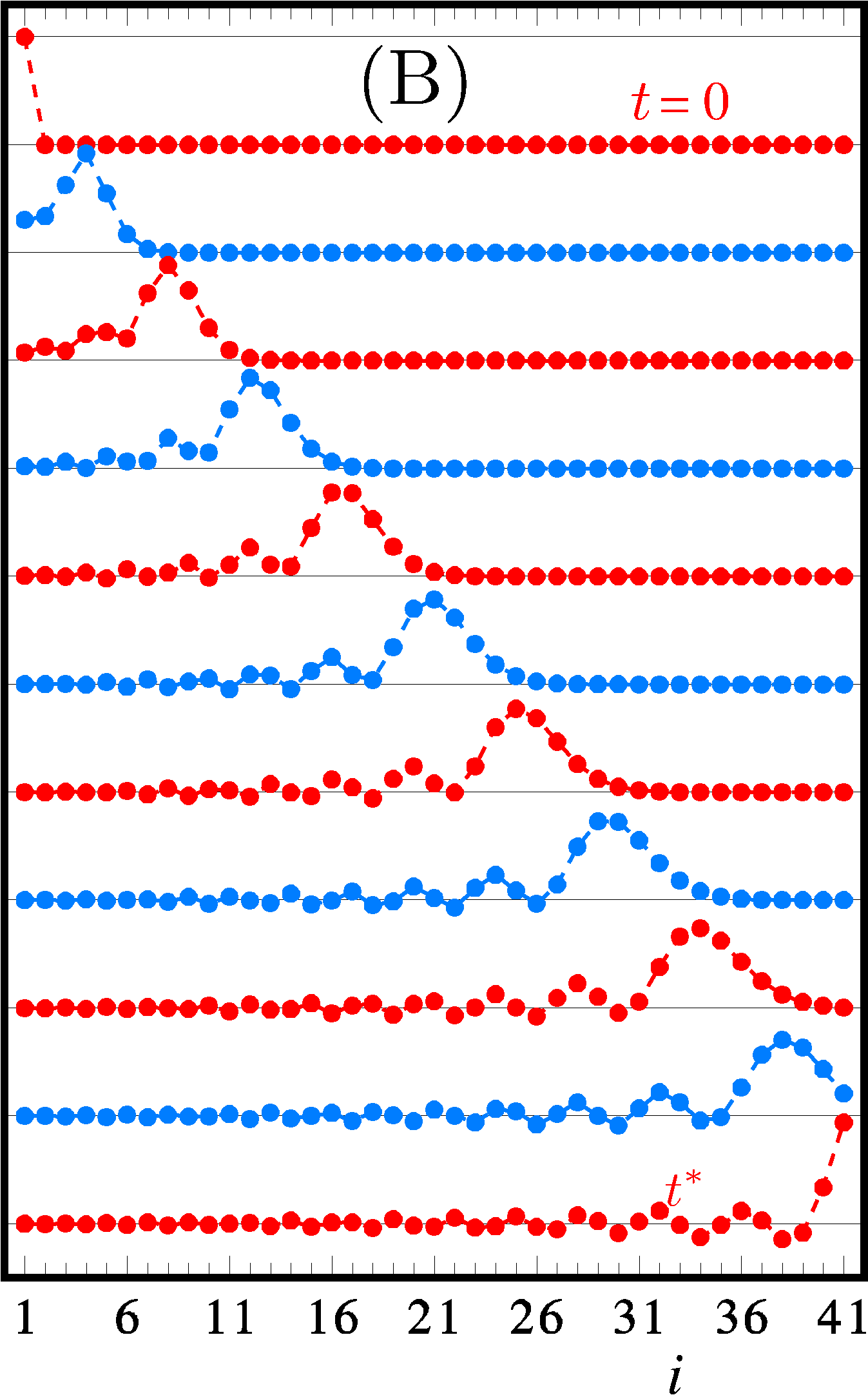}\\[3mm]
\includegraphics[width=0.4\textwidth]{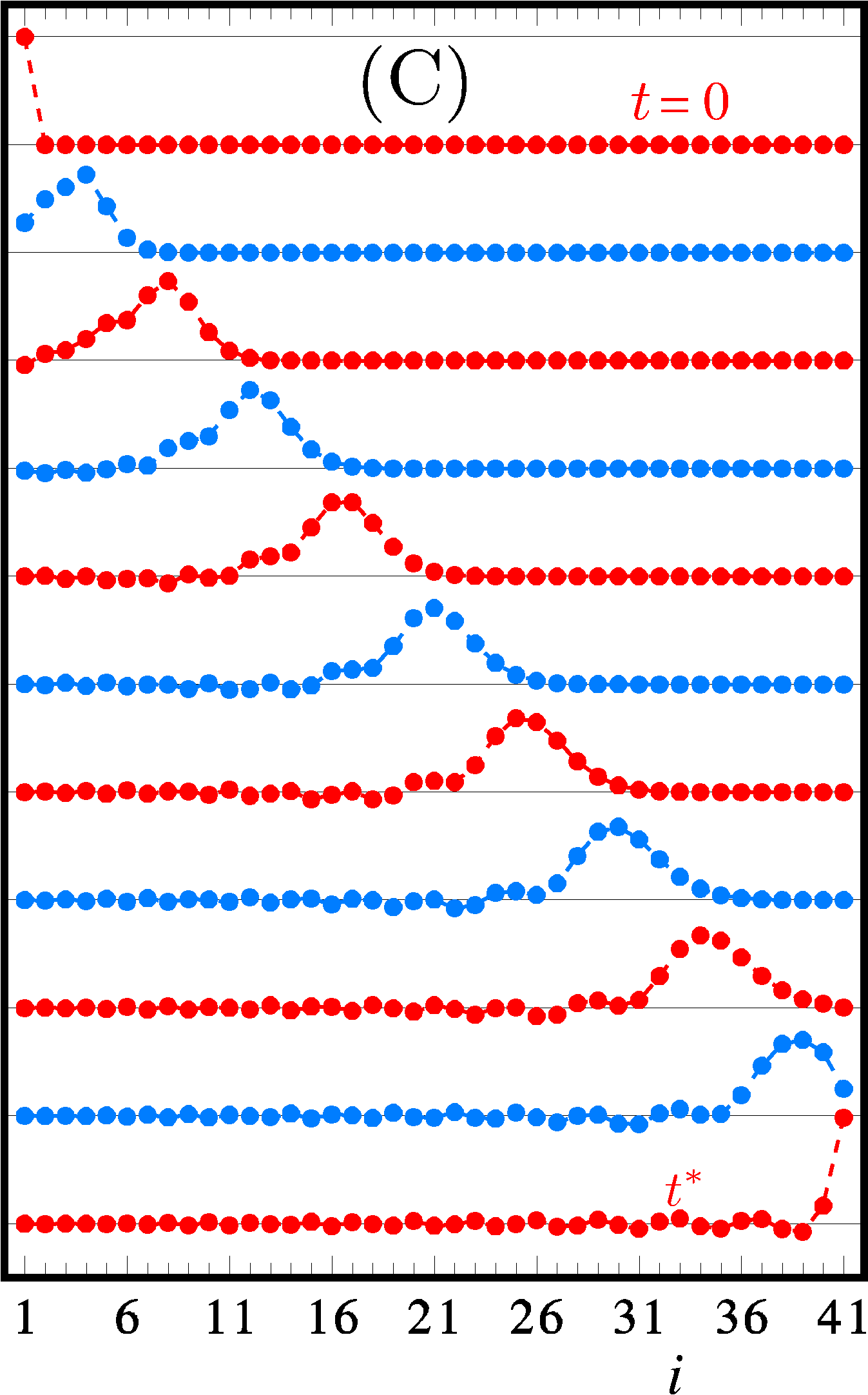}~~~~~
\includegraphics[width=0.4\textwidth]{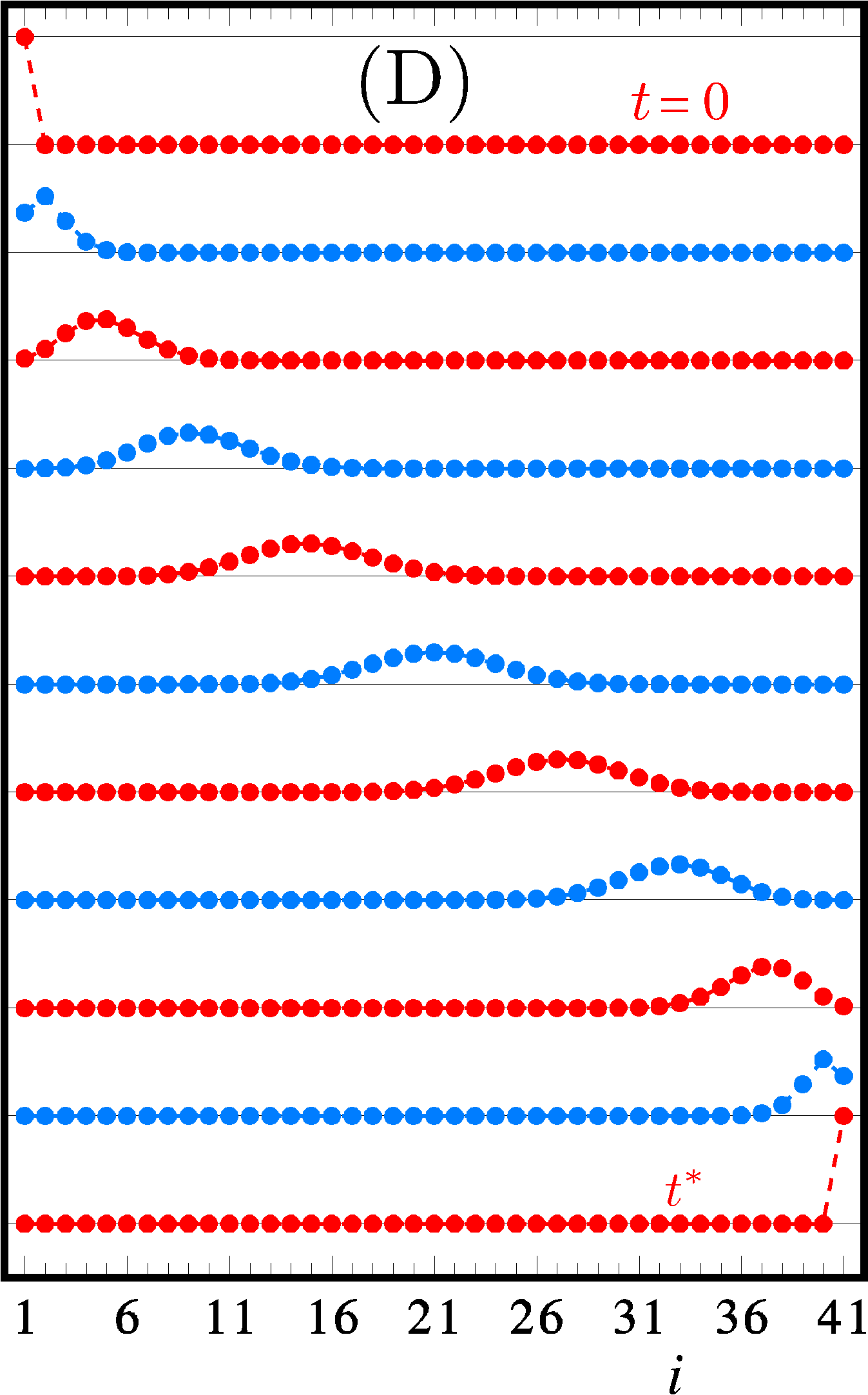}
\caption{$N=41$ chain, snapshots of the dynamics at equal time intervals of $t^*/10$ between $0$ and $t^*$. The first three panels correspond to the columns of Table~\ref{t.N41n}, namely the uniform chain (A), the quasi-uniform chains with optimal extremal mass $m_1$ (B) and with 2 optimized extremal masses, $m_1,~m_2$, and their spring $K_1$ (C); the last panel is the perfect chain (D) reported in Table~\ref{t.N41p}.  Animations are available as Supplemental Material~\cite{supplmat}.
}
	\label{f.N41}
\end{figure}

\begin{figure}
\includegraphics[width=0.4\textwidth]{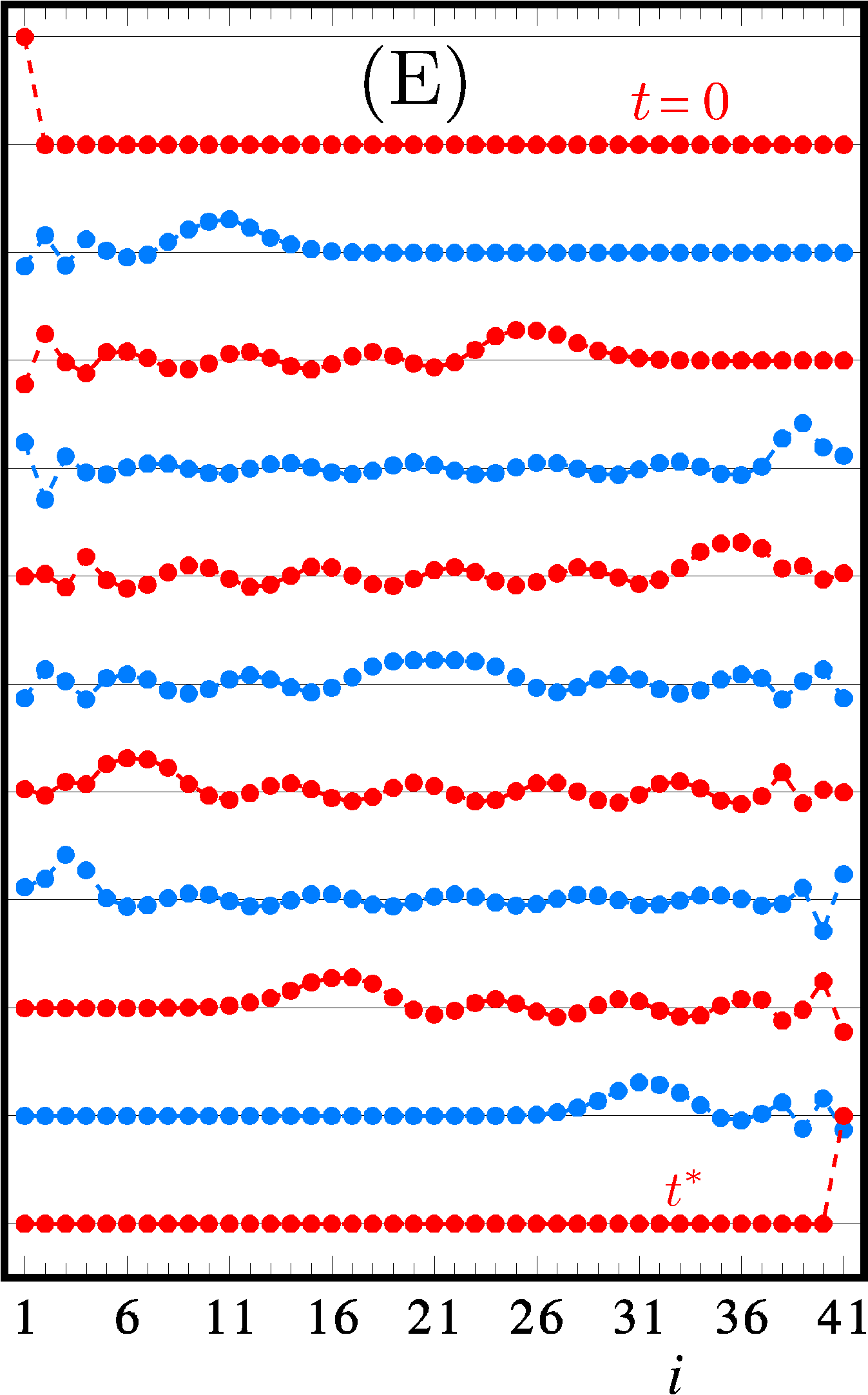}~~~~~
\includegraphics[width=0.4\textwidth]{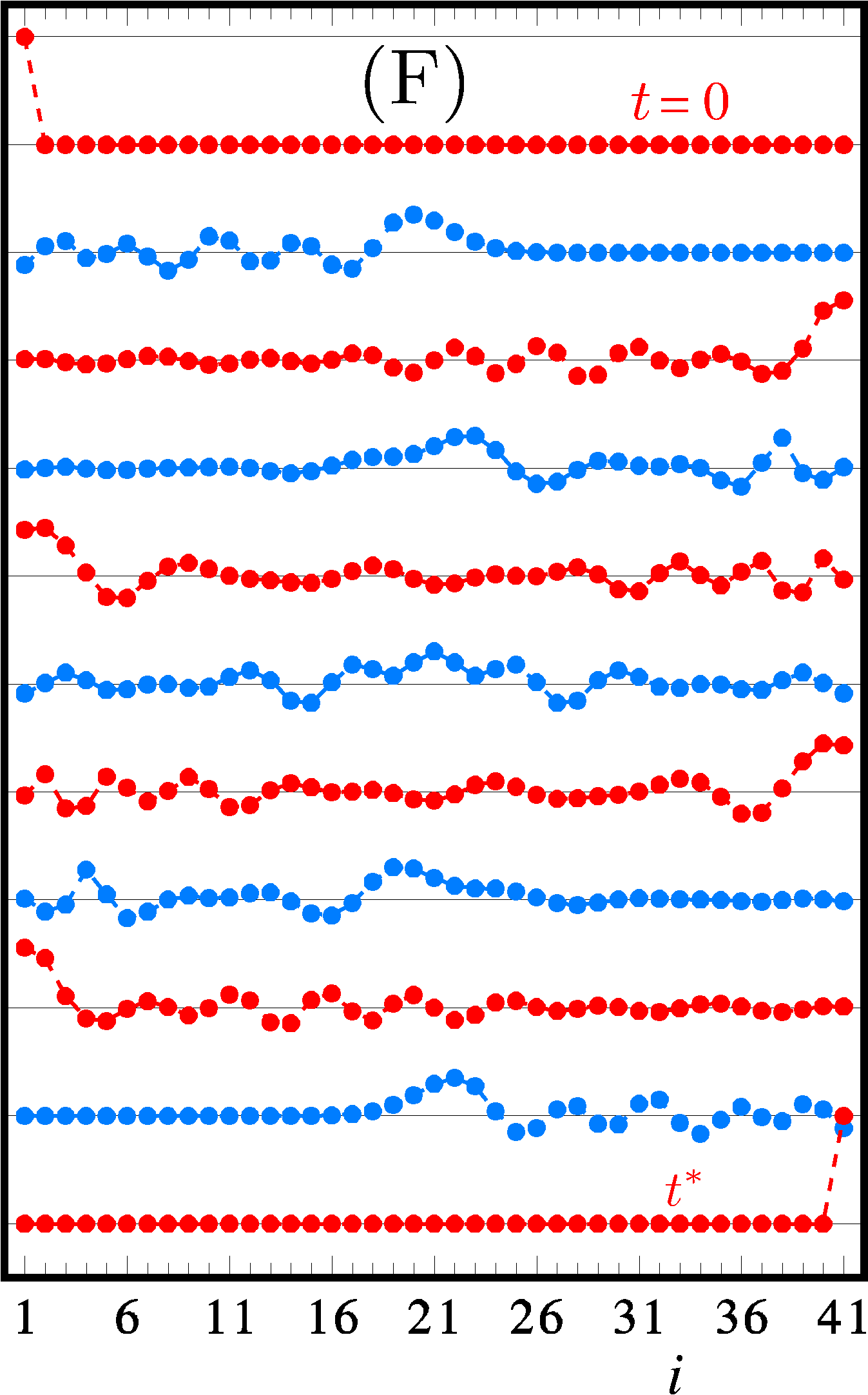}
\caption{Snapshots of the dynamics at equal time intervals of $t^*/10$ between $0$ and $t^*=10$ for the two alternative $N=41$ perfectly transmitting chains (E) and (F) described in Table~\ref{t.N41p}.  Animations are available as Supplemental Material~\cite{supplmat}. }
	\label{f.N41p}
\end{figure}


\begin{thebibliography}{00}
\bibitem{note2}
It is to be noted that periodicity occurs for any set of integers $\{k_n\}$, without the parity constraint.
\end{thebibliography}
\end{document}